\documentclass[namedreferences]{SolarPhysics}
\usepackage[optionalrh,solaenum]{spr-sola-addons} % For Solar Physics 
\usepackage{graphicx}        % For eps figures, newer & more powerfull
\usepackage{color}           % For color text: \color command
\usepackage{url}             % For breaking URLs easily trough lines
\begin{document}

\begin{article}

\begin{opening}

\title{Multiwavelength study of a Solar Eruption from AR NOAA 11112: II. Large-Scale Coronal Wave and Loop Oscillation}
\author{Pankaj~\surname{Kumar}$^{1}$\sep
        K.-S.~\surname{Cho}$^{2,3,1}$\sep
        P. F.~\surname{Chen}$^{4}$\sep
        S.-C.~\surname{Bong}$^{1}$\sep
       Sung-Hong~\surname{Park}$^{1}$\sep	
       }
\runningauthor{P. Kumar et al.}
\runningtitle{Large-scale Coronal Waves and Loop Oscillations}
\institute{$^{1}$ Korea Astronomy and Space Science Institute (KASI), Daejeon,
	305-348, Republic of Korea.  email: \url{pankaj@kasi.re.kr}}
\institute{$^{2}$ NASA Goddard Space Flight Center, Greenbelt, Maryland, USA.}
\institute{$^{3}$ Department of Physics, The Catholic University of America,
	Washington, D. C., USA.}
\institute{$^{4}$ School of Astronomy and Space Science, Nanjing University,
	Nanjing 210093, China.}
%%%%%%%%%%%%%%%%%%%%%%%%%%%%%%%%%%%%%%%%%%%%%%%%%%%%%%%%%%%%%%%%%%%%%%%%
\begin{abstract}
 We analyse multiwavelength observations of an M2.9/1N flare that occurred
in AR NOAA 11112 on 16 October 2010. 
AIA 211 \AA \ EUV images reveal the presence of a faster coronal wave
(decelerating from $\approx$1390 to $\approx$830 km s$^{-1}$) propagating ahead of a slower wave (decelerating from $\approx$416 to $\approx$166 km s$^{-1}$) towards the western limb.
The dynamic radio spectrum from Sagamore Hill
radio telescope shows the presence of
metric type II radio burst, which reveals the presence of a coronal shock
wave (speed$\approx$800 km s$^{-1}$). The speed of the faster coronal wave
derived from AIA 211 \AA \ images is found to be comparable to the coronal shock
speed. AIA 171 \AA \ high-cadence observations showed that a coronal loop,
which was located at the distance of $\approx$$0.32 R_\odot$ to the west of the
flaring region, started to oscillate by the end of the impulsive phase of the
flare. The results indicate that the faster coronal wave may be the first
driver of the transversal oscillations of coronal loop. As the slower wave
passed through the coronal loop, the oscillations became even stronger. There was a plasmoid eruption observed in EUV and a white-light CME was recorded, having velocity of $\approx$340-350 km s$^{-1}$.
STEREO 195 \AA \ images show an EIT wave, propagating in the same direction of the lower-speed coronal wave observed in
AIA, but decelerating from $\approx$320 to $\approx$254 km s$^{-1}$. These observations reveal the co-existence of both waves ({\it i.e.} coronal
Moreton and EIT waves), and type II radio burst seems to be associated with the
coronal Moreton wave.  

\end{abstract}
\keywords{Solar flare -- coronal loops, magnetic field, flux rope, magnetic reconnection.}
\end{opening}
%-------------------------------------------------

%%%%%%%%%%%%%%%%%%%%%%%%%%%%%%%%%%%%%%%%%%%%%%%%%%%%
\section{Introduction}
Large-scale coronal waves are often observed during solar eruptions. For
example, the so-called EIT waves are the transient wavelike disturbances in the solar
corona that propagate with the typical speed of 170--350 km s$^{-1}$ followed
 by the expanding coronal dimming \cite{thompson1998,thompson1999,klassen2000}.
 These were first observed by EUV imaging Telescope (EIT) onboard SOHO 
\cite{del1995}. It is now widely accepted that EIT waves are associated with
coronal mass ejections (CMEs) rather than solar flares 
\cite{dela99,biesecker2002,cliver2005,chen2006}. 
Regarding the spatial relationship between EIT waves and CMEs, some authors
found that they are cospatial \cite{chen2009,dai2010}, whereas some others
claimed that EIT wave fronts are ahead of the CME leading edge
\cite{veronig2008,pats2009a,pats2009b,kien2009,kien2011,veronig2010,muhr2011}. EIT waves
were usually explained as the fast-mode magnetoacoustic waves in the corona
({\it e.g.,} \opencite{wang00}, \opencite{wu01}), therefore they would be the coronal counterparts of
the H$\alpha$ Moreton waves that are observed in the chromosphere with a velocity
of $\approx$500-2000 km s$^{-1}$ \cite{moreton1960}. The fast-mode wave model
was first questioned by \inlinecite{dela99}. Furthermore, \inlinecite{eto2002}
investigated CME-related waves in an X-class flare event and found that EIT
wave front is not cospatial with the Moreton wave front inferred from filament
winking, and the propagation speeds of both waves were clearly different.
Therefore, several non-wave models were later developed see (\opencite{will09}; \opencite{warmuth2010}; \opencite{gall11}; \opencite{chen11};
\opencite{zhuk11} for reviews). On the basis of MHD numerical
simulation, \inlinecite{chen2002} proposed that EIT waves are apparently
moving brightenings, which are generated by the successive stretching of the
closed field lines pushed by an erupting flux rope. According to the 
field-line stretching model \cite{chen2002,chen2005}, a fast-mode
magnetoacoustic wave (or coronal Moreton wave) should be ahead of the EIT wave
in a CME event, which was confirmed by \inlinecite{harr03}. Recently, using
the high-resolution SDO/AIA observations, \inlinecite{chen2011} convincingly
reported the existence of the fast-mode coronal Moreton wave ({\it i.e.} coronal counterpart of Moreton wave), which is three
times faster than the EIT wave. 

Whereas EIT waves show a good correlation with the decimetric type II radio
bursts, the speed derived from the type II radio burst is usually three times
larger than the EIT wave speed \cite{klassen2000}. The speed of Moreton wave,
however, matches with the speed derived from type II radio bursts 
\cite{eto2002,warmuth2004}. This suggests that the Moreton wave, rather than the EIT wave,
and the type-II radio burst are two aspects of a single phenomenon, or the MHD
fast-mode shock propagating in the corona \cite{uchida1974}. Furthermore, note that there is a significant fraction of events where EIT waves are found to be a coronal signature directly associated with Moreton waves, {\it i.e.,} both are 
cospatial \cite{warmuth2001,warmuth2004,warmuth2005,vrsnak2002,veronig2006,muhr2010}. 

Besides the debates on EIT waves, the origin of coronal shock waves (usually evident in the form of type II radio burst) is also under debate (for detail please see, \opencite{warmuth2007}, \opencite{chen2011a}). It may be driven by two possible physical mechanisms, {\it i.e.} (i) a blast wave ignited by the pressure pulse of a flare \cite{vrsnak1995,vrsnak2000a,vrsnak2000b,khan2002,narukage2002,hudson2003,mag2010}, (ii) a piston-driven shock due to a CME \cite{klassen1999,klassen2003,cho2011}. Thus, coronal shock waves may be associated with solar flares, CMEs, or some combination of these phenomena \cite{magara00,mag2008,vrsnak2008}. 

In this paper, we analyse the multiwavelength observations from SDO/AIA and
STEREO to investigate the a large-scale coronal wave event and its impact on
the solar corona in terms of loop oscillations. In Section 2, we will present
multiwavelength observations of the large-scale coronal waves and the CME. In
the last section, we will discuss the results and draw conclusions.  

%*****************************************************************************
%***************************************************************************

%%%%%%%%%%%%%%%%%%%%%%%%%%%%%%%%%%%%%%%%%%%%%%%%%%%%%%%%%%%%%%%%
\section{Observations}

 %%%%%%%%%%%%%%%%%%%%%%%%%%%%%%%%%%%%%%%%%%%%%%%%%%%%%%%%%%%%%%%%%%%%%%%%%%%%%%
\begin{figure}
\centerline{
\includegraphics[width=1.0\textwidth]{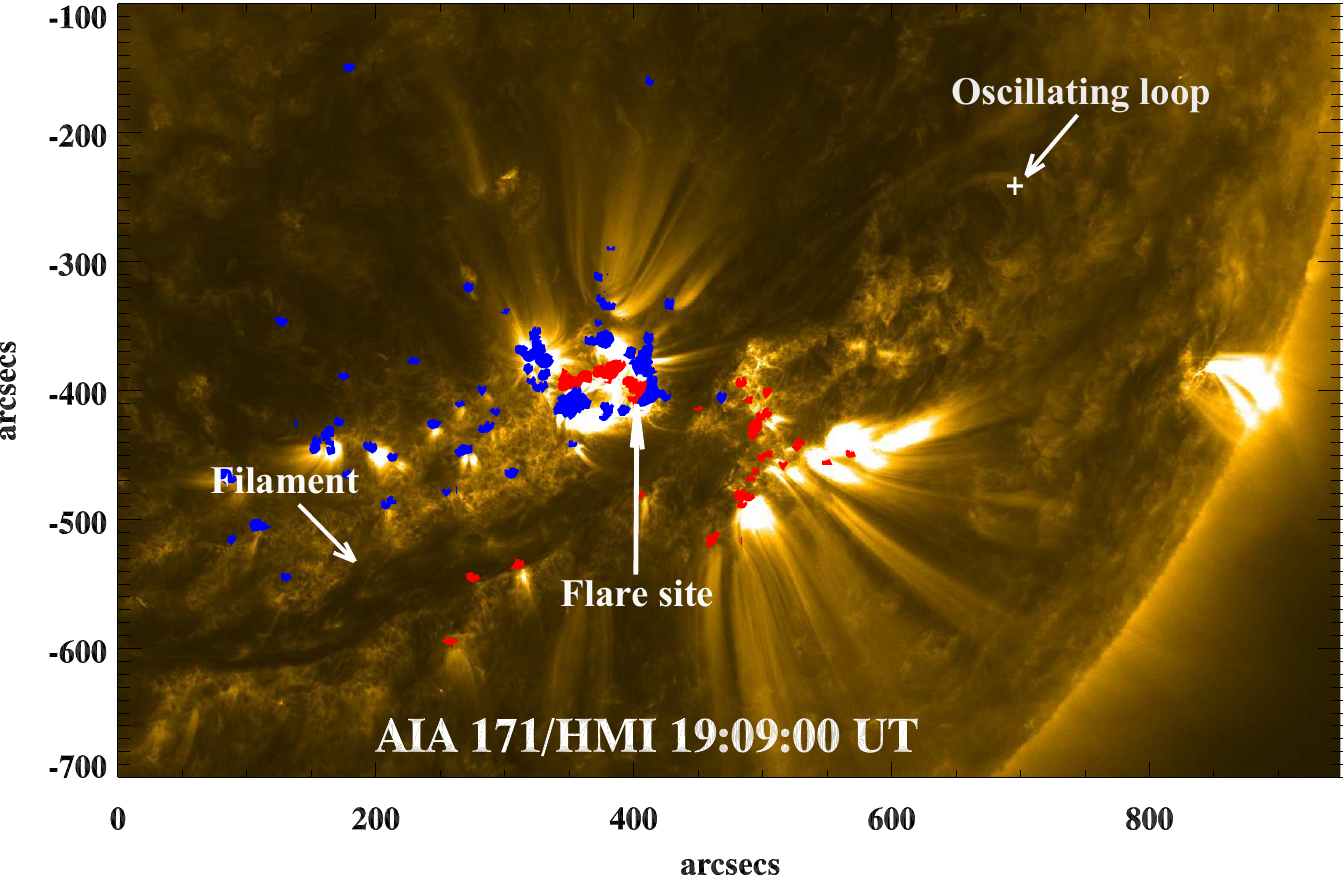}
}
\caption{SDO/AIA 171 \AA \ image overlaid by HMI magnetogram contours (red=positive, blue=negative) showing a huge filament system lying along the polarity inversion line, and the sites of the flare and the oscillating loop system. The loop apex is marked by the `+' symbol.}
\label{aia171}
\end{figure}
%%%%%%%%%%%%%%%%%%%%%%%%%%%%%%%%%%%%%%%%%%%
\begin{figure}
\centerline{
\includegraphics[width=0.5\textwidth]{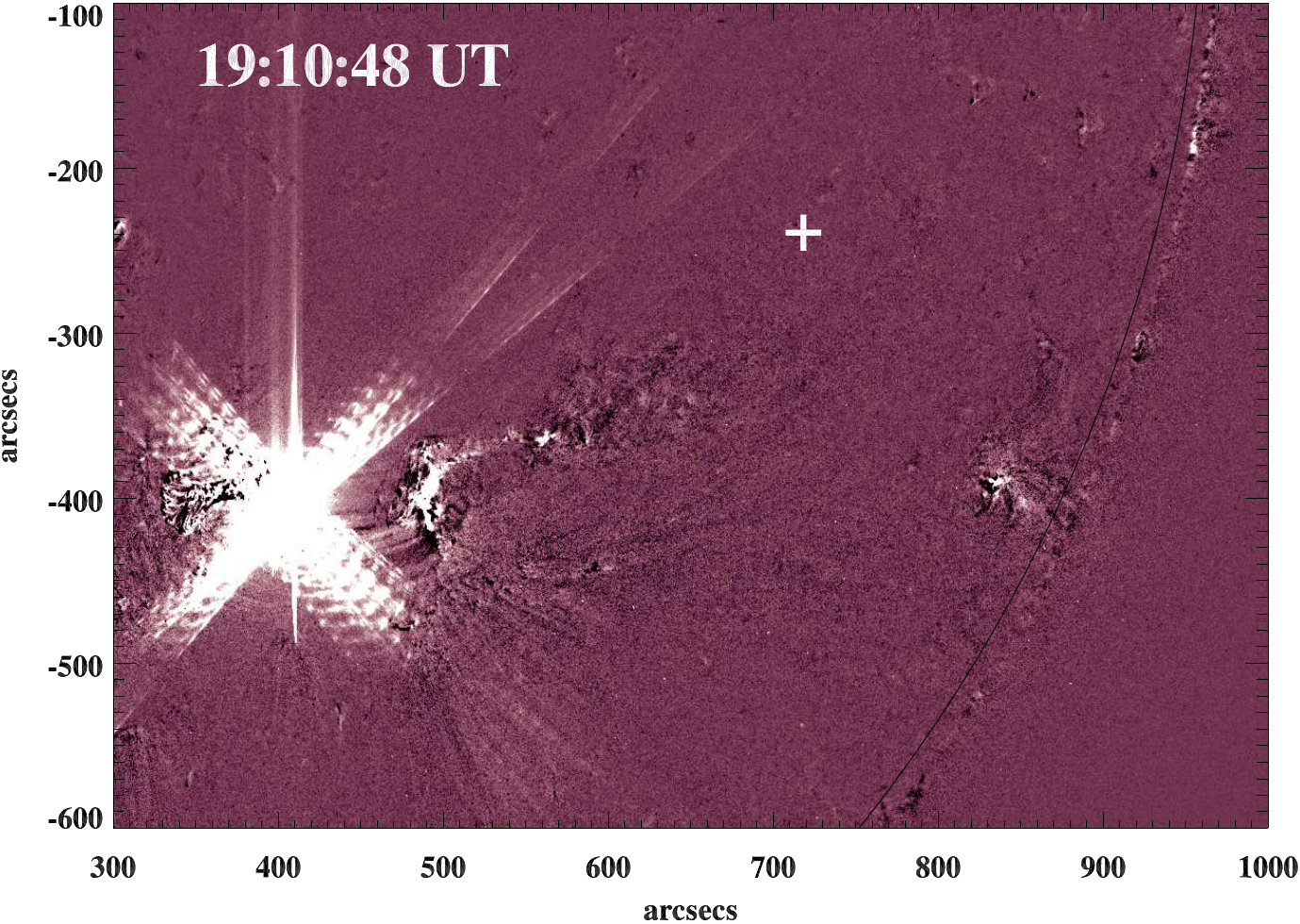}
\includegraphics[width=0.5\textwidth]{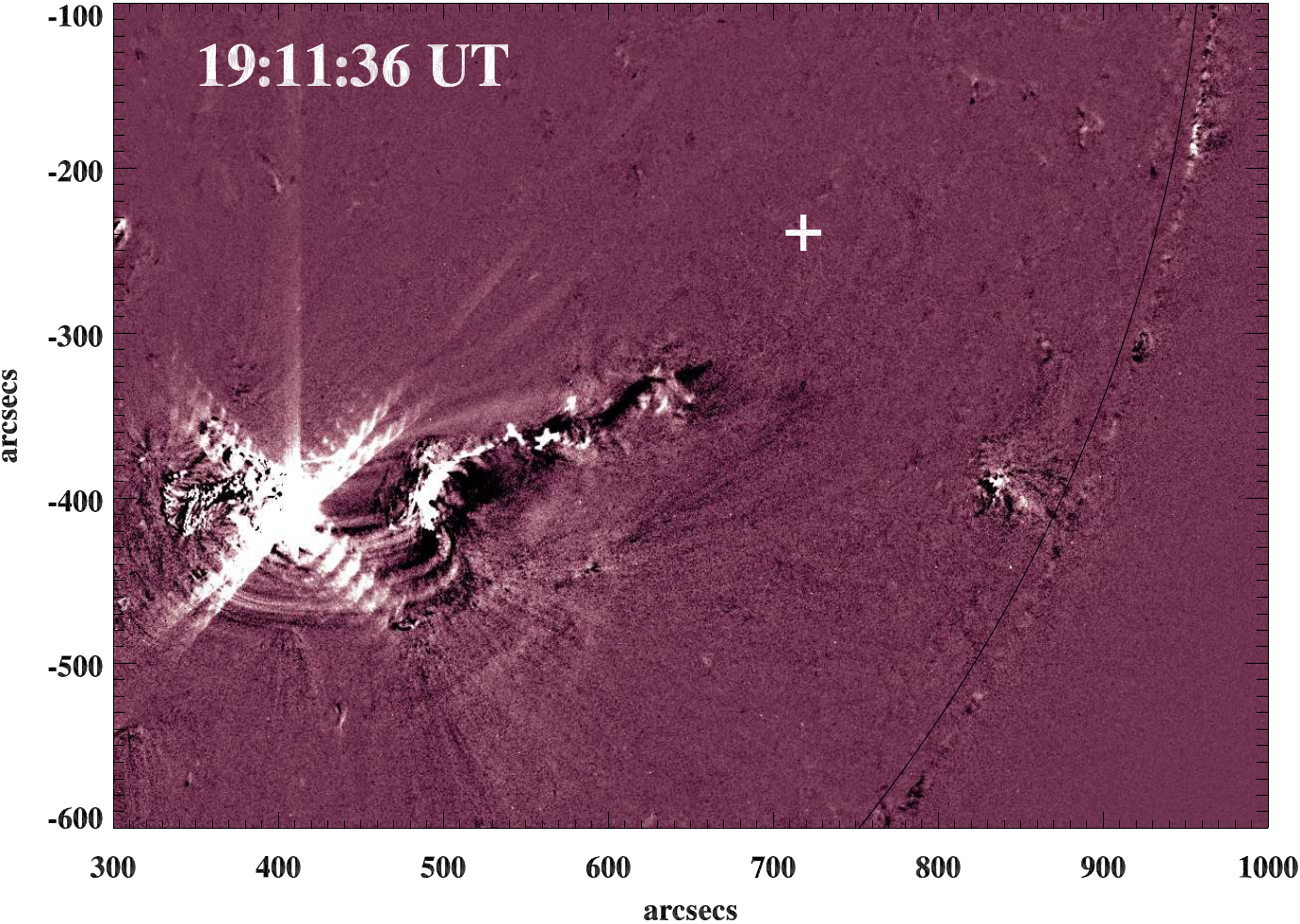}
}
\centerline{
\includegraphics[width=0.5\textwidth]{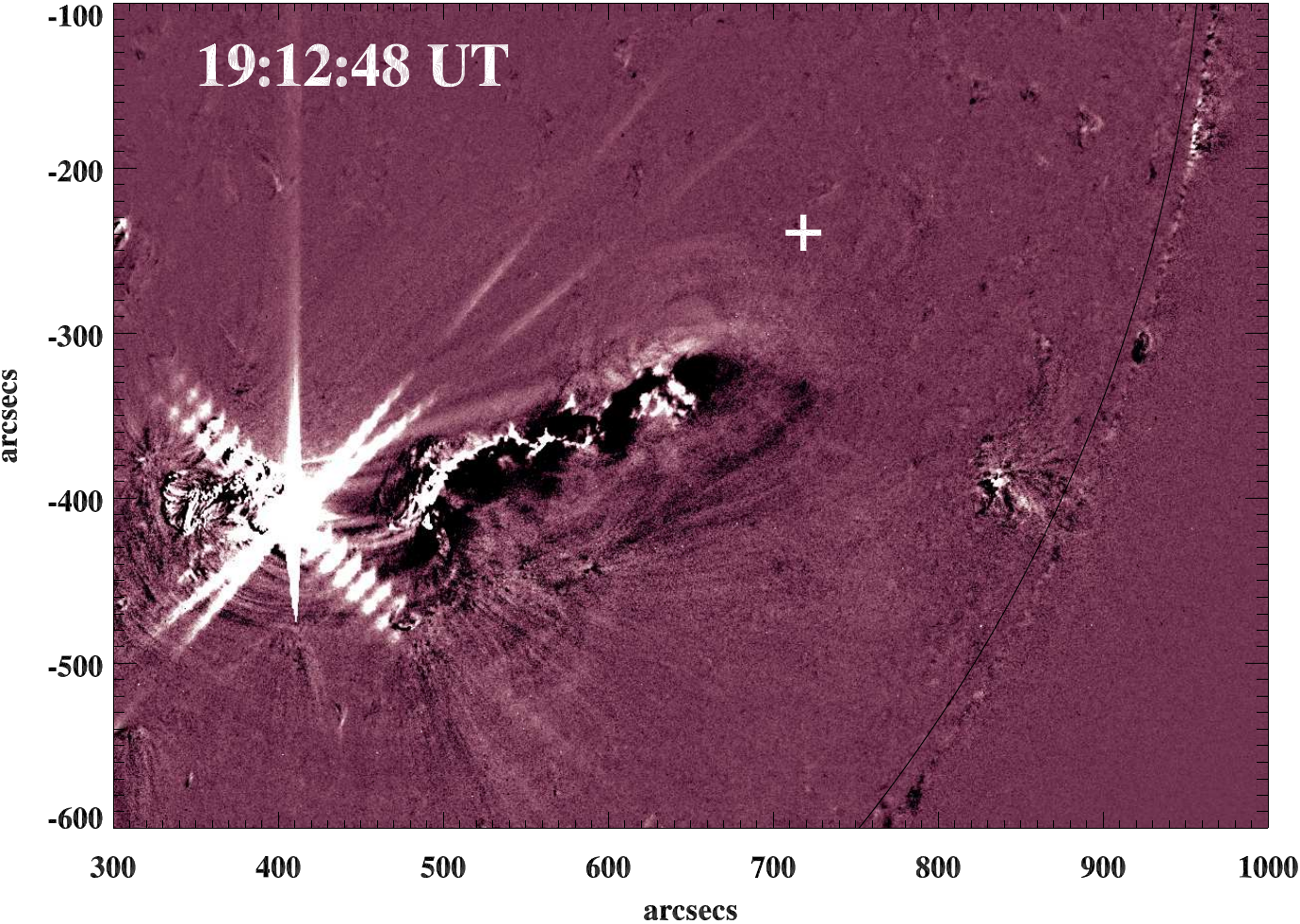}
\includegraphics[width=0.5\textwidth]{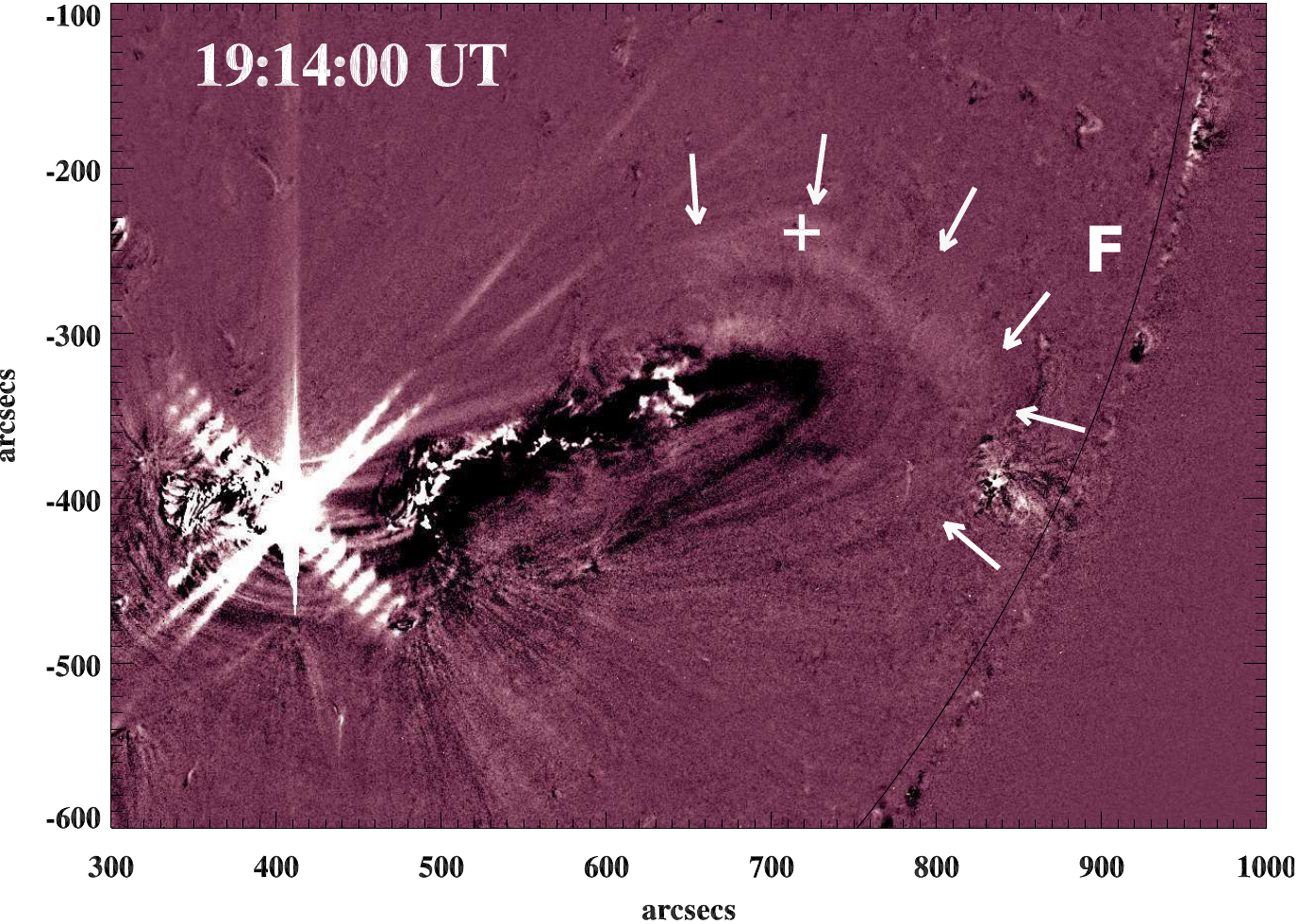}
}
\centerline{
\includegraphics[width=0.5\textwidth]{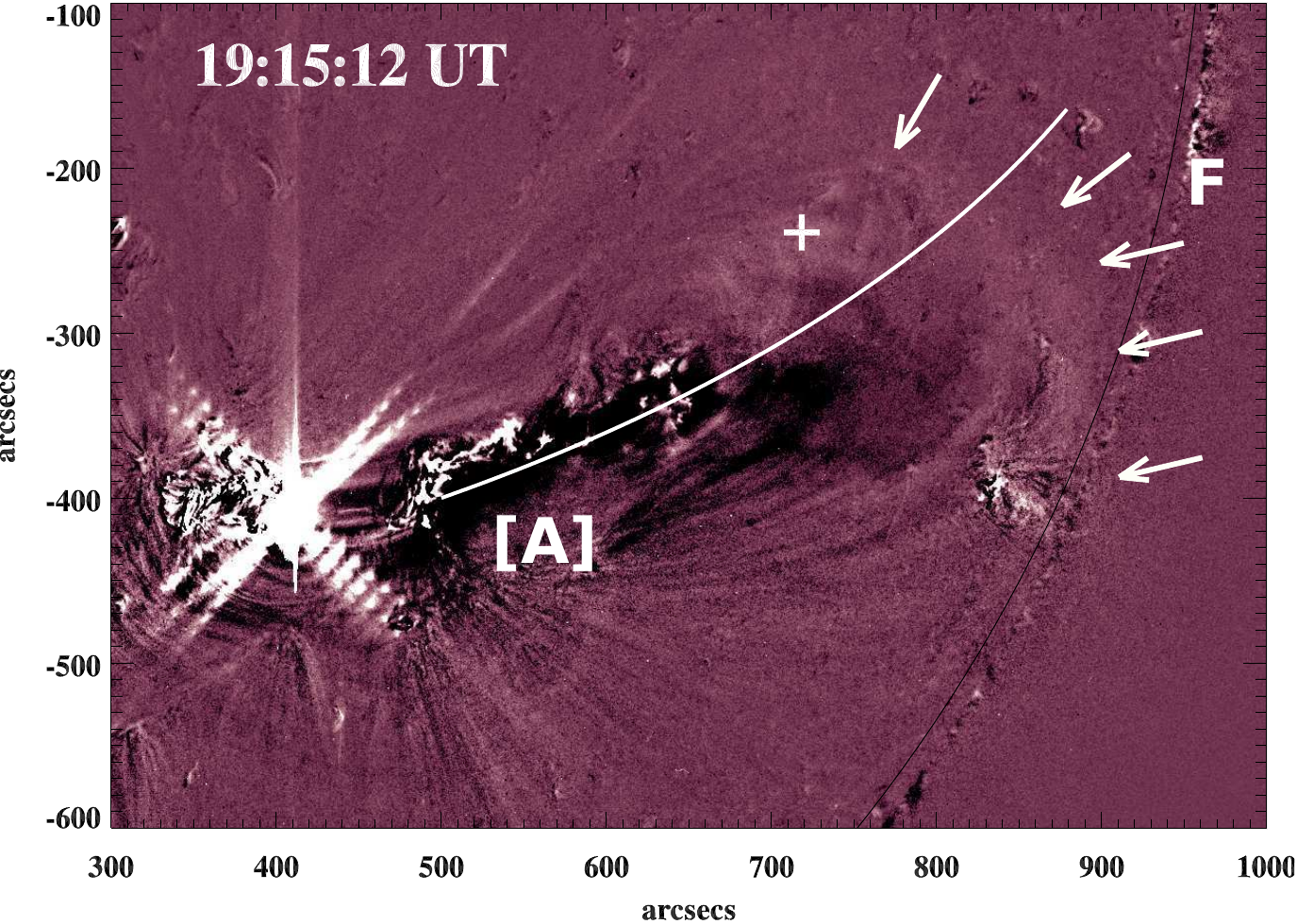}
\includegraphics[width=0.5\textwidth]{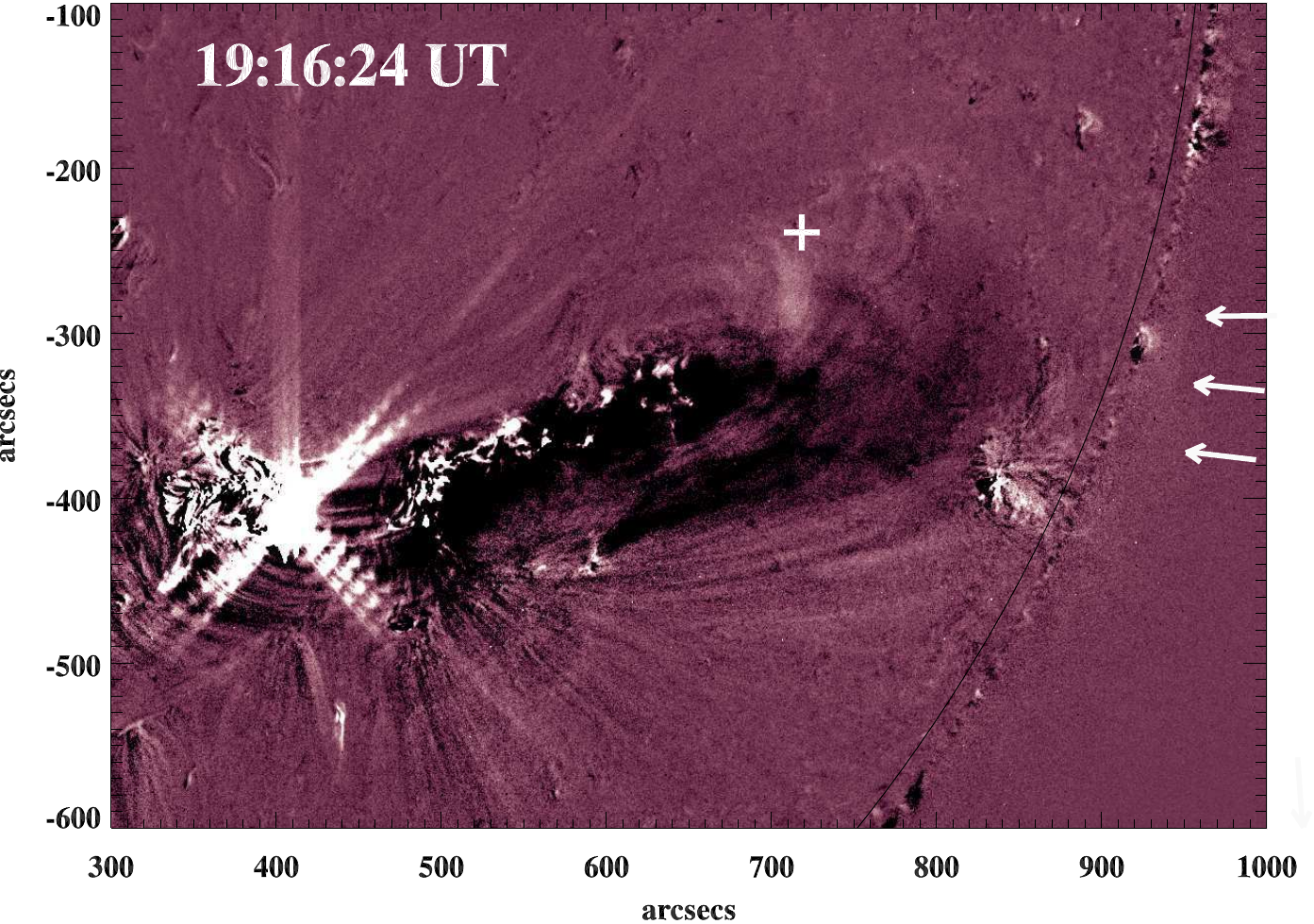}
}
\centerline{
\includegraphics[width=0.5\textwidth]{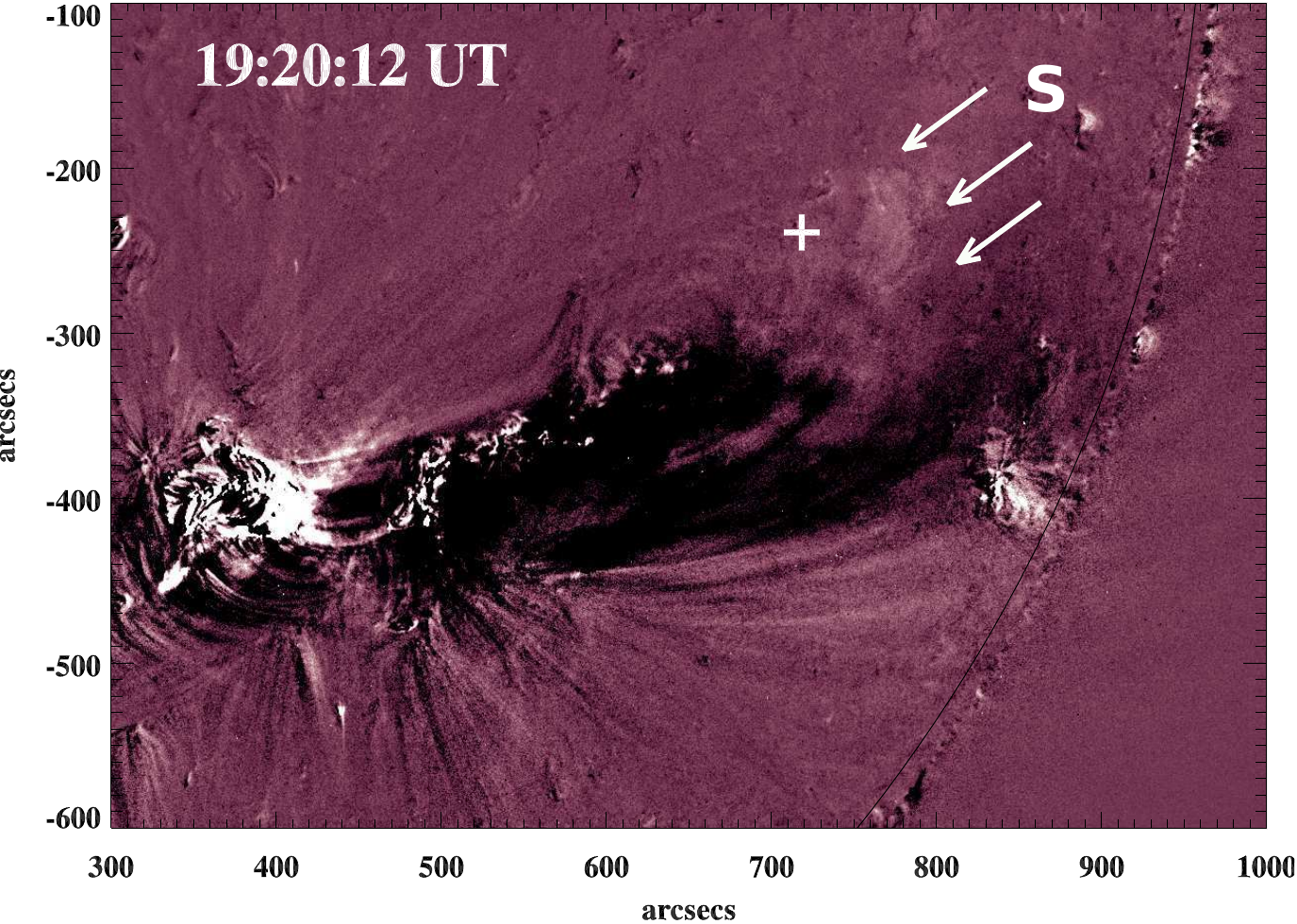}
\includegraphics[width=0.5\textwidth]{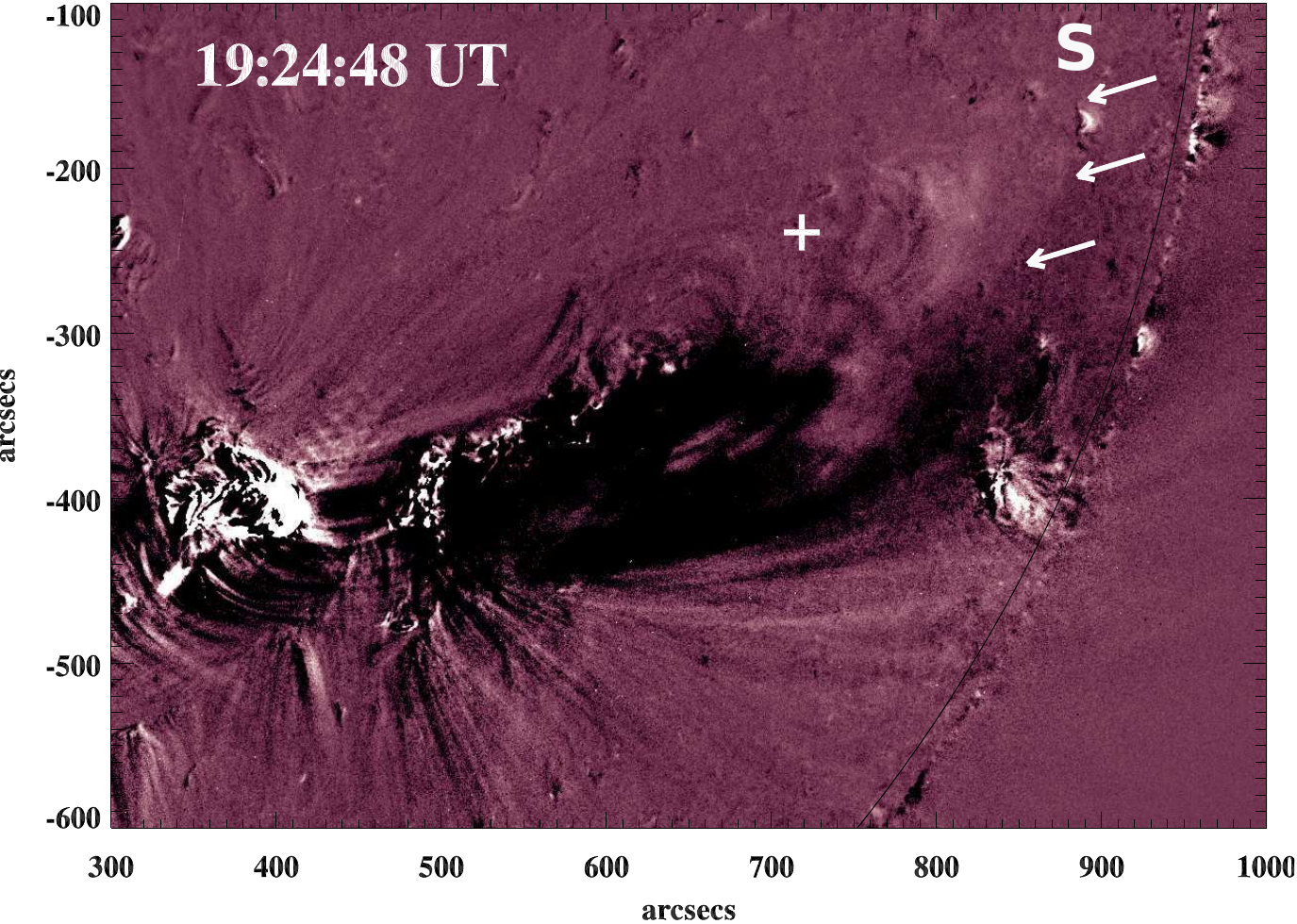}
}
\caption{SDO/AIA EUV 211 \AA \ base-difference images showing the propagation of coronal waves (indicated by arrows). The location of the oscillating coronal loop apex is indicated by the `+' symbol in each image. The loop started oscillating when the leading edge of the faster wave approached the loop system. White line `A' shows the great circle along the solar surface in the direction of wave propagation. The faster and slower waves are indicated by `F' and `S', respectively.}
\label{aia211}
\end{figure}             
%*****************************************************************************
\begin{figure}
\centerline{
\includegraphics[width=0.5\textwidth]{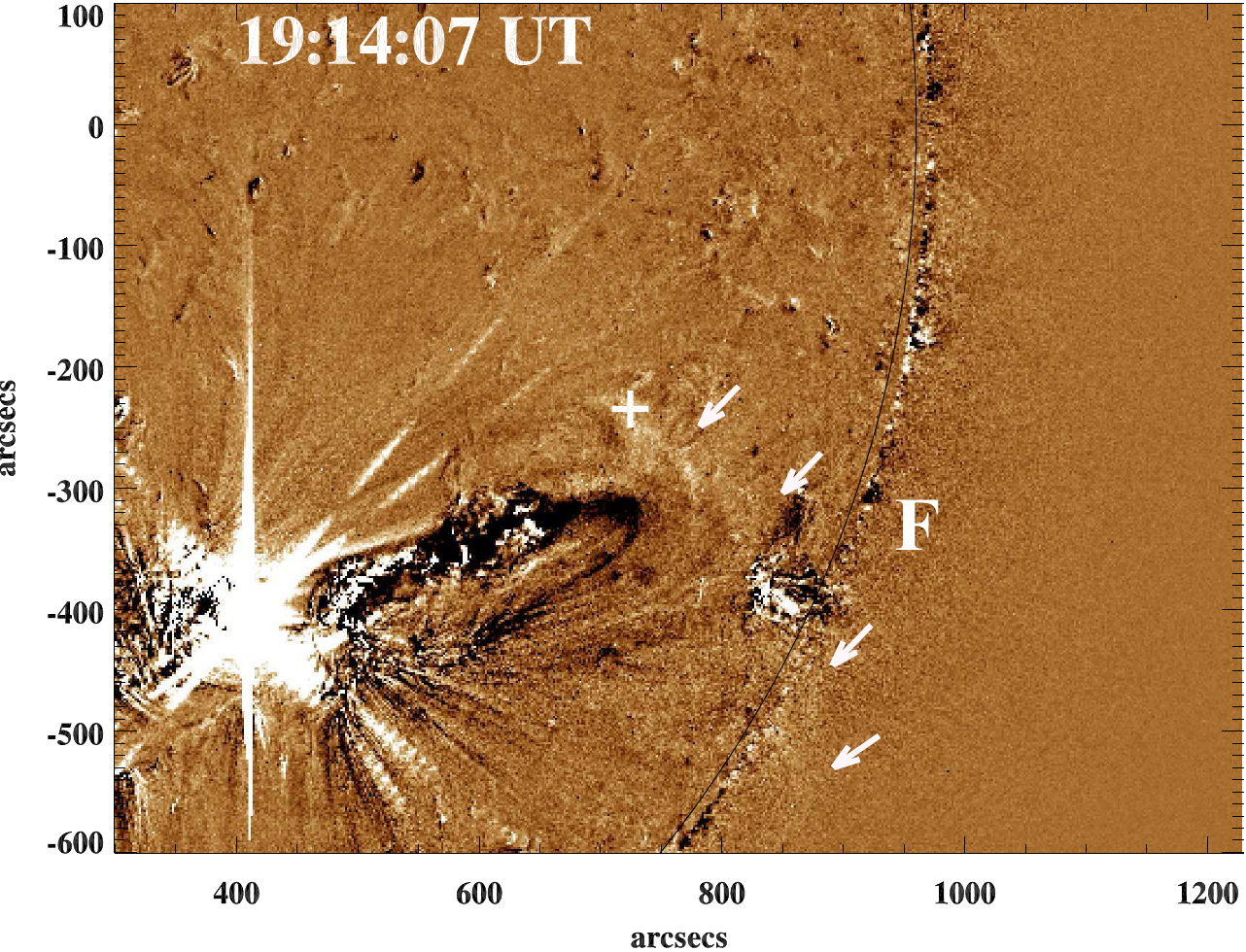}
\includegraphics[width=0.5\textwidth]{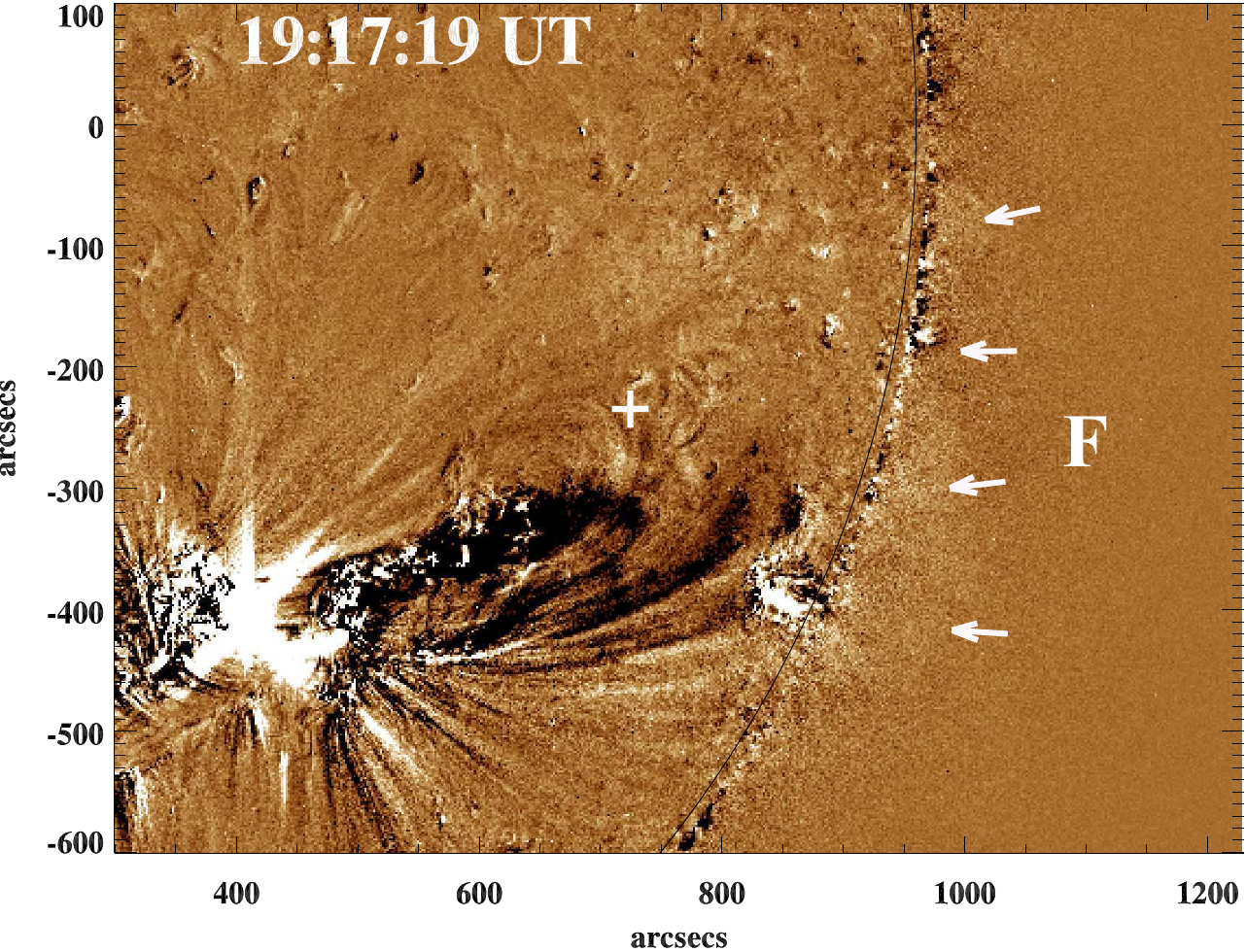}
}
\centerline{
\includegraphics[width=0.5\textwidth]{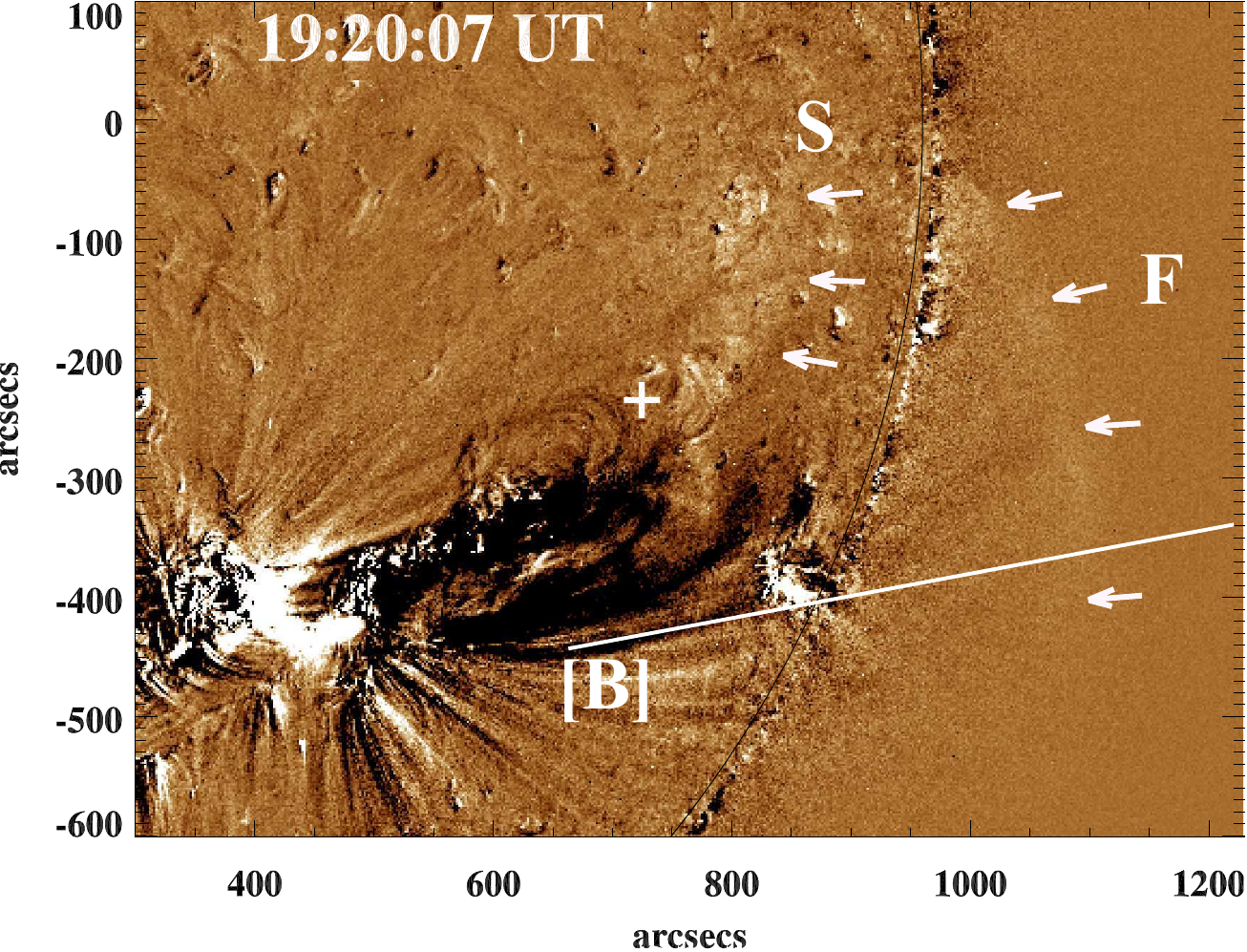}
\includegraphics[width=0.5\textwidth]{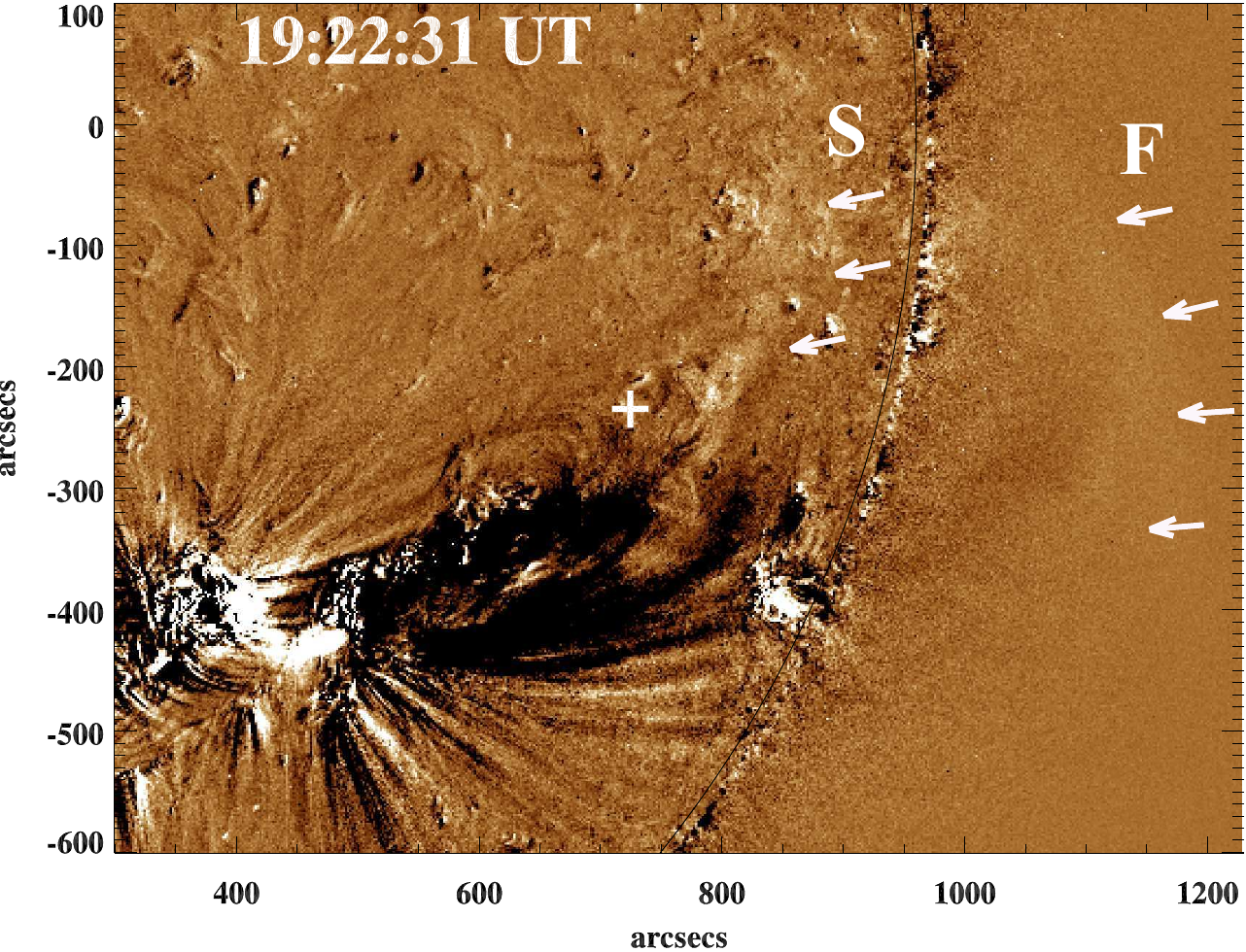}
}
\caption{SDO/AIA EUV 193 \AA \ base-difference images showing the propagation of coronal waves (indicated by the arrows). The location of the oscillating coronal loop apex is indicated by the `+' symbol in each image.  The line indicated by `B'  shows the slice cut along the direction of wave propagation. The faster and slower waves are indicated by `F' and `S', respectively.}
\label{aia193}
\end{figure}             
%*****************************************************************************

%*****************************************************************************
\begin{figure}
\centerline{
\includegraphics[width=0.8\textwidth]{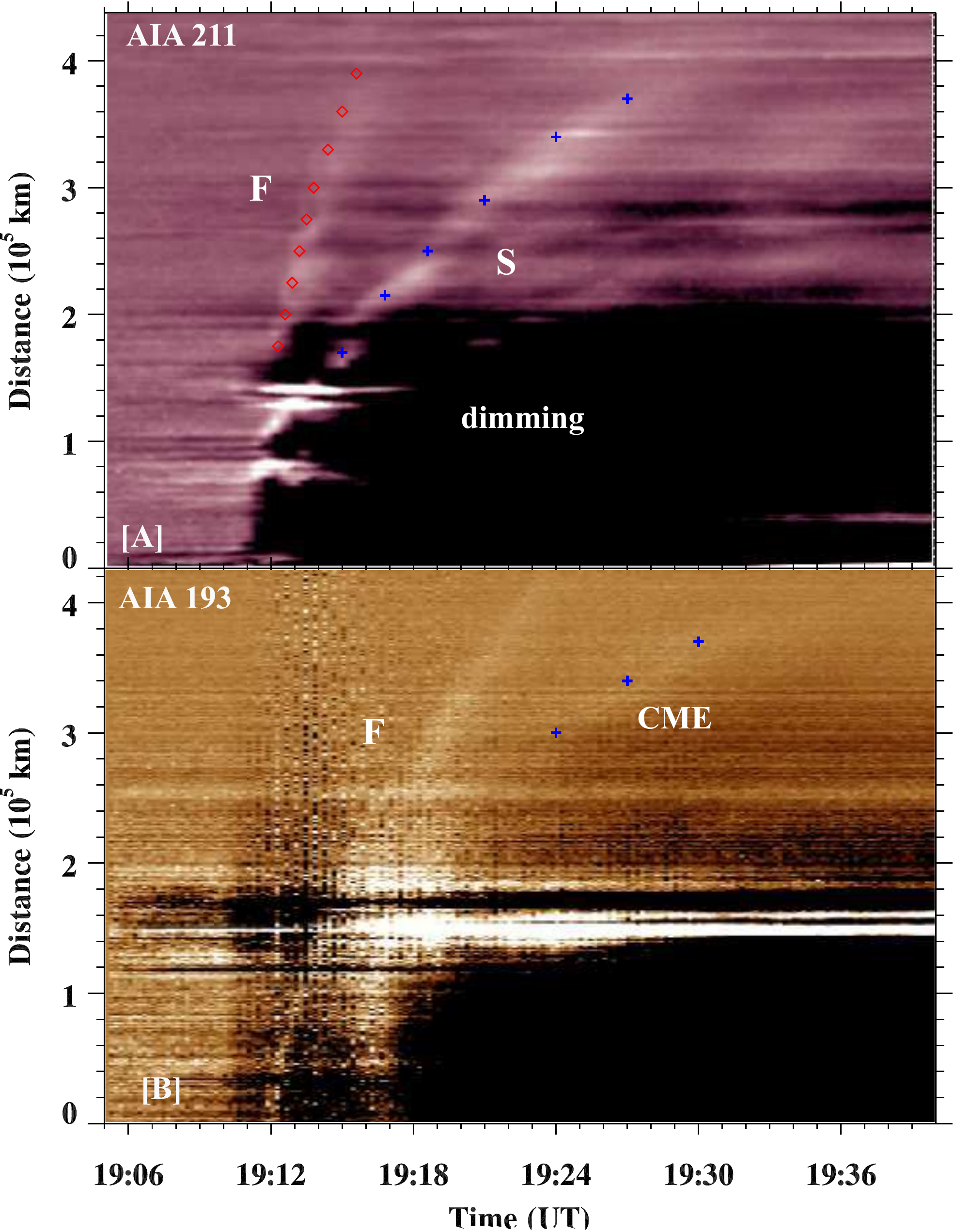}
}
\caption{Time evolutions of the 211 \AA \ and 193 \AA \ base-difference
intensity distributions along slice `A' (top) and slice `B' (bottom), respectively. Faster and slower waves are marked by `F' and `S',
respectively. The mean speeds (from the linear fit) of the faster and slower waves
are respectively $\approx$1086 km s$^{-1}$ and $\approx$276 km s$^{-1}$.}
\label{st}
\end{figure}
%********************************************************************************
%*************************************************************************
\begin{figure}
\centerline{
\includegraphics[width=0.25\textwidth]{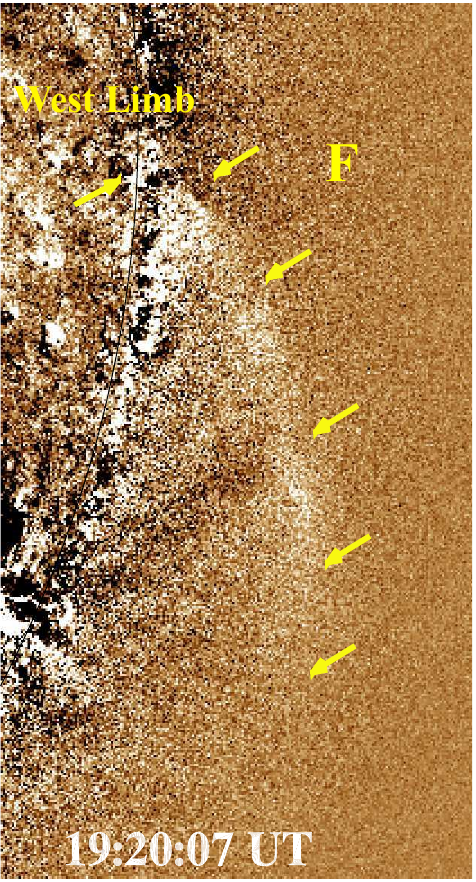}
\includegraphics[width=0.25\textwidth]{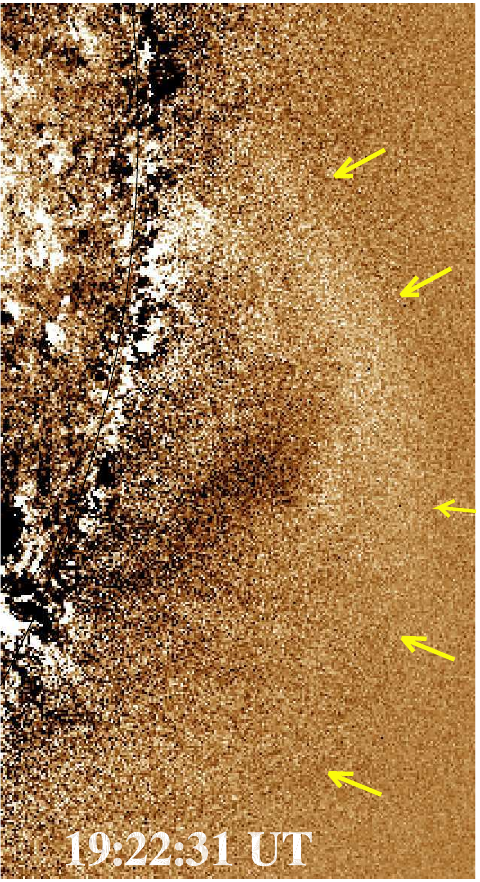}
\includegraphics[width=0.25\textwidth]{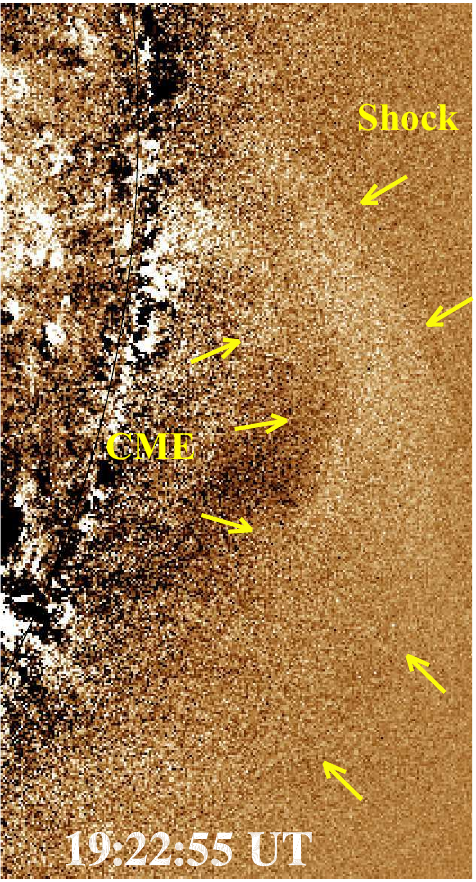}
\includegraphics[width=0.25\textwidth]{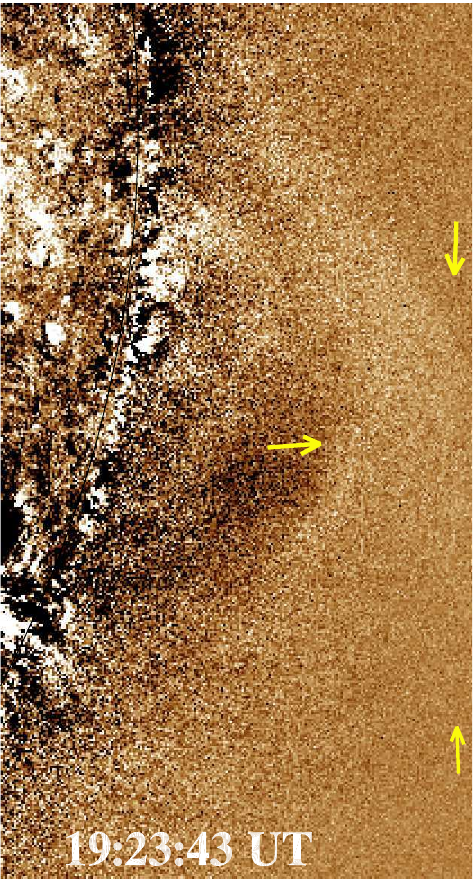}
}
\centerline{
\includegraphics[width=0.25\textwidth]{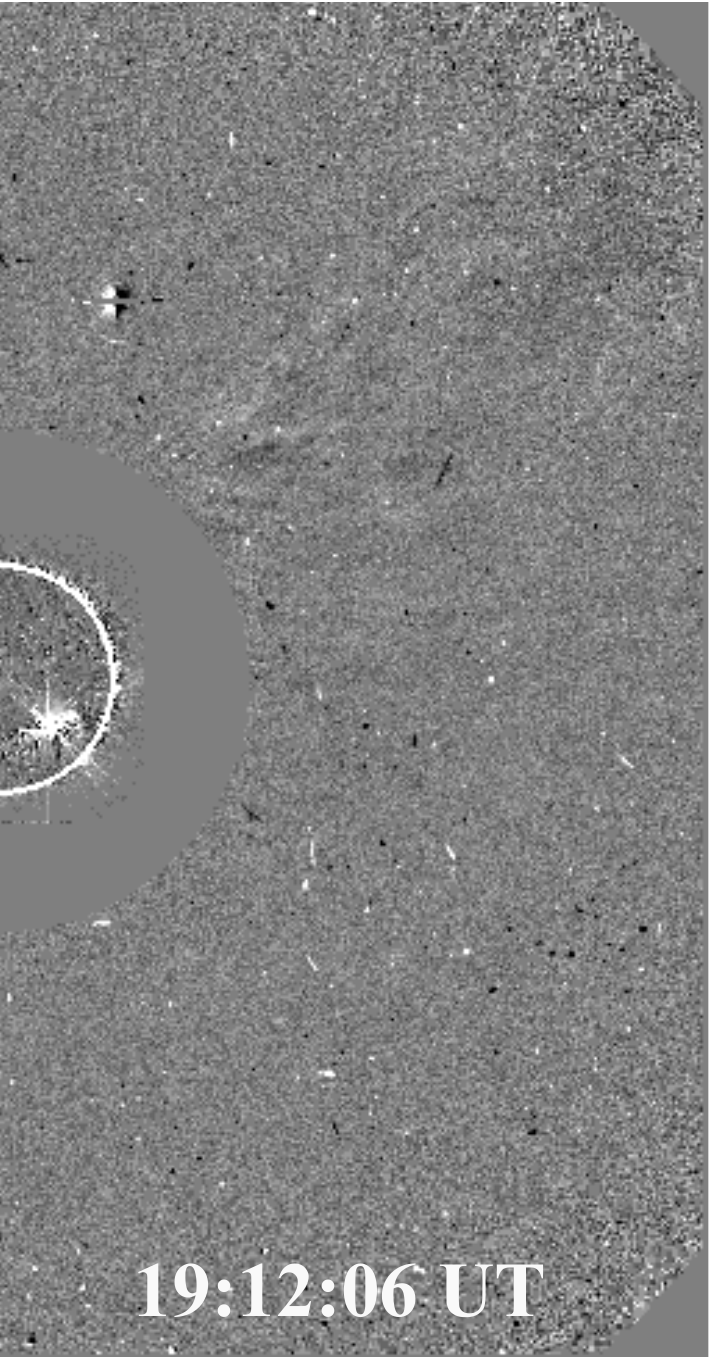}
\includegraphics[width=0.25\textwidth]{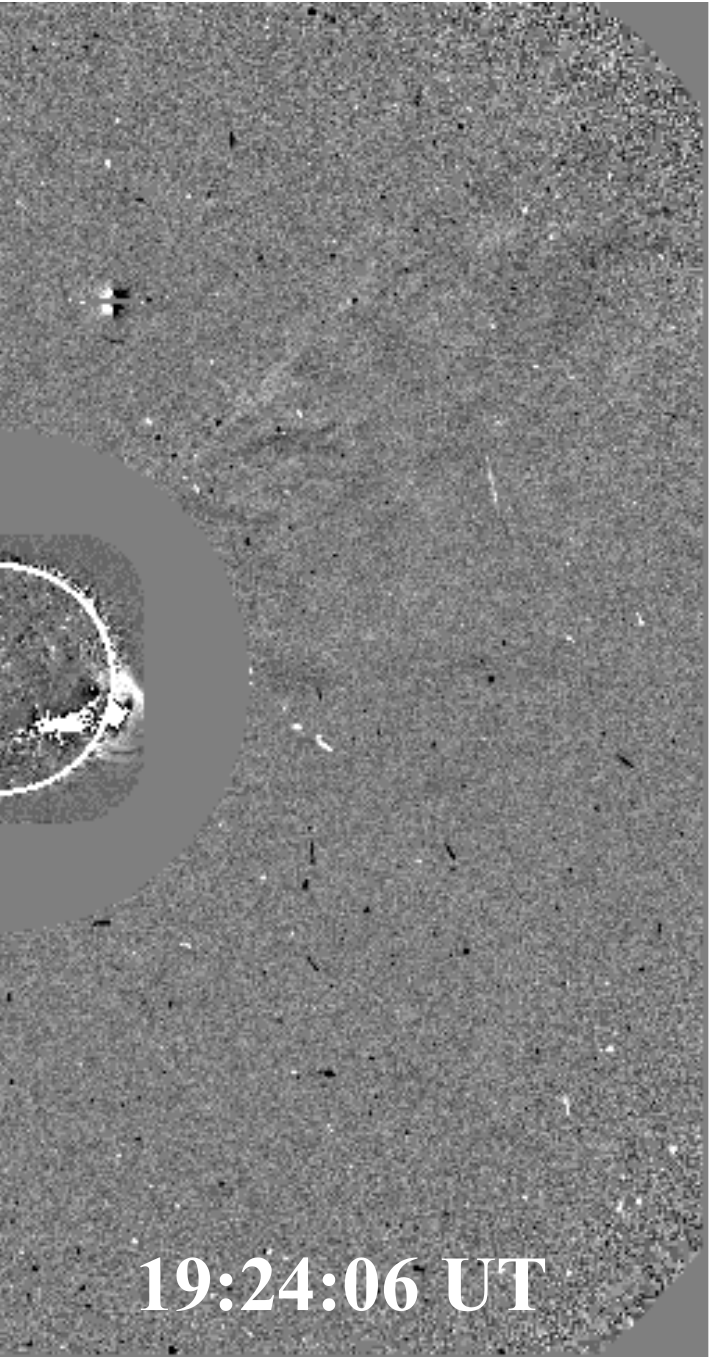}
\includegraphics[width=0.25\textwidth]{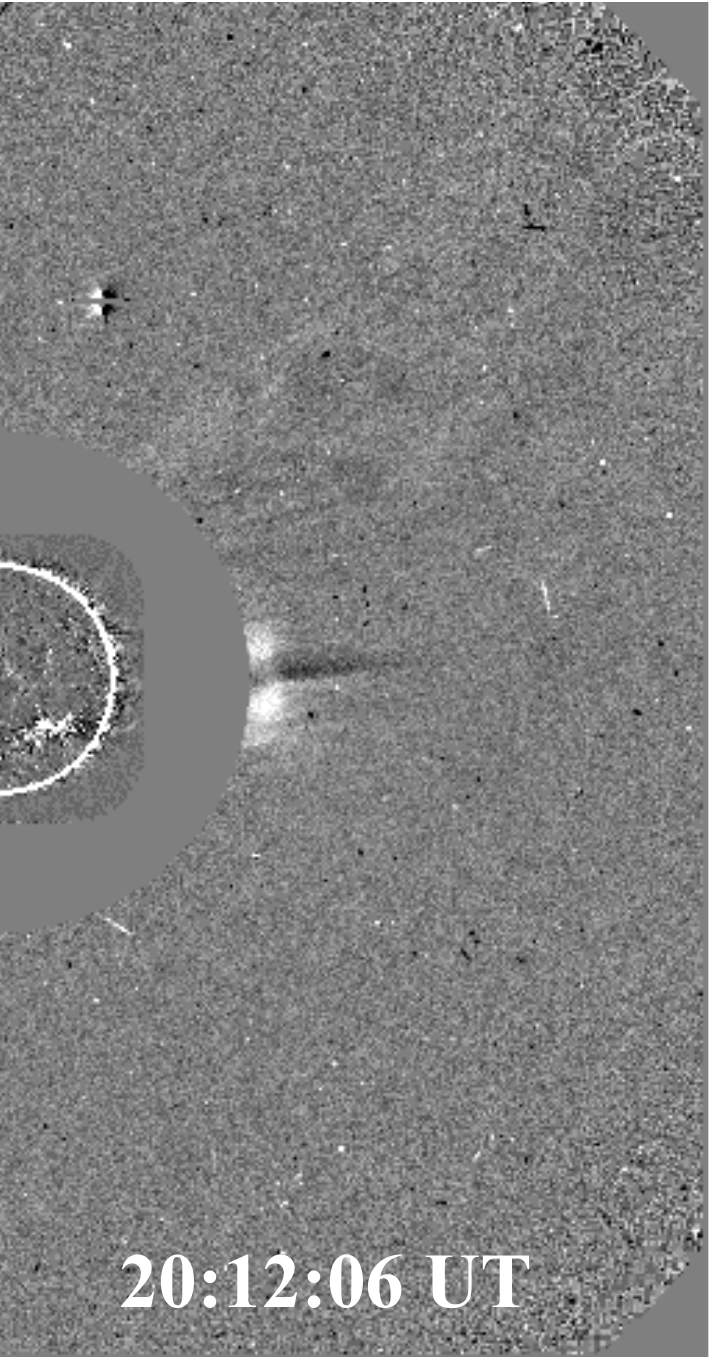}
\includegraphics[width=0.25\textwidth]{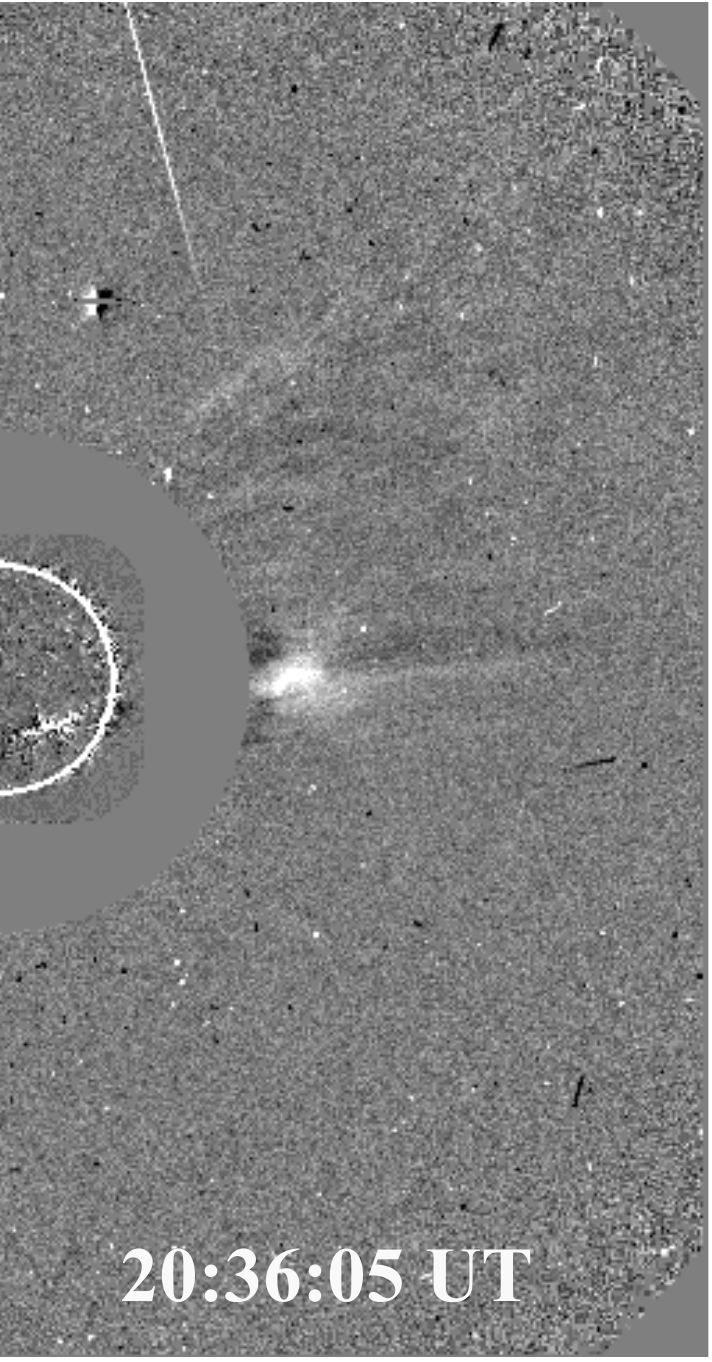}
}
\caption{Top panel: AIA 193 \AA \ base-difference images showing the propagating faster wavefront `F' ahead of expanding CME loop. The size of each image is 380$^{\prime\prime}$$\times$700$^{\prime\prime}$.
Bottom panel: AIA 193 \AA \ and LASCO C2 composite difference images showing the CME associated with the flare with a speed$\approx$350 km s$^{-1}$.}
\label{lasco}
\end{figure}

%*************************************************************************
 The \textit{Atmospheric Imaging Assembly} (AIA: \opencite{lemen2012}) on board the Solar Dynamics
Observatory (SDO: \opencite{pesnell2012}) mission provides multiple high-resolution full-disk images
of the corona and transition region. The field of view of
each image is $1.3R_\odot$. The pixel resolution of the images is 
0.6$^{\prime\prime}$ and the cadence is 12 s. We use AIA 171, 211 and 193 
\AA\ EUV observations to investigate the evolution of coronal waves
associated with an M2.9 flare that occurred in AR NOAA 11112 on 16 October
2010. The detailed description of flare energy build up and triggering mechanism has been discussed in \inlinecite{kumar2012}(hereafter Paper I). This paper consists of the description of flare/CME associated coronal waves kinematics as well as the interaction of these waves with the coronal loop, which showed transverse oscillations during the passage of the waves through it.

\subsection{Coronal Waves}
Figure \ref{aia171} displays SDO/AIA 171 \AA\ EUV image overlaid by SDO/HMI
magnetogram contours to show the magnetic environment in a larger field of
view. Red/blue represents positive/negative polarity field region. We can see a huge filament lying along the
polarity inversion line (PIL), which did not erupt during the flare. The flare site is indicated by an
arrow. A small loop system, indicated by another arrow, was located $\approx$$0.32R_{\odot}$
away from the flare site on its west side. \inlinecite{asc2011} have studied
extensively the properties of the transversal oscillations of this loop system.
They have interpreted it as a kink mode transversal oscillations and studied the
properties of the MHD modes and diagnosed the local plasma conditions of the
oscillating loop system. In particular, they noticed that, unlike most
previously studied events, the oscillation of this coronal loop showed no
damping for several periods.

In this paper, we only investigate the most probable driver of loop
oscillations under the baseline of multiwavelength observations of the M-class
flare and the associated large-scale wave phenomena. We use
base-difference images to reduce the artifacts and for the correct information
about the waves \cite{attrill2010}. For investigating the driver of loop
oscillations, we make AIA 211 \AA \ base-difference images. The selected
base-difference images are displayed in Figure \ref{aia211}. AIA 211 \AA \ EUV
images are sensitive to the temperature of 2 MK. The `+' symbol marks the
location of the small coronal loop in each image. The first image at 19:10:48
UT shows the flare site as well as the extended bright flare ribbon towards
the west direction. We can see the propagating disturbance/wave towards west
along the direction of bright ribbons (19:11:36 UT). Coronal dimmings were
observed behind the propagating wavefront probably due to the depletion in
plasma density. At 19:14:00 UT, the nearly circular shape of the fast
wavefront is evident in the image, which is indicated by the arrows and marked
by `F'. At this time it approached the site of coronal loop system indicated
by the `+' symbol (shown in 171 \AA\ image, Figure \ref{aia171}), which
started to oscillate. The `F' front continued to expand towards west in a
ballooning shape and it could be tracked close to the western limb (shown by
arrows at 19:16:24 UT). In the meanwhile, we see another bright wavefront at
19:16:24 UT behind the faster front, which was also approaching the loop site.
This is a slow wavefront, indicated by `S' (19:20:12 UT) and it slowly passed
through the loop site (see images at 19:20:12 and 19:24:48 UT).  Therefore,
these images reveal the existence of both faster and slower coronal waves
which propagated towards the west side of the flare site.

Figure \ref{aia193} displays the selected base-difference images during the
flare. We plot a larger field of view in these images in order to show the
propagation of the faster wave from the solar disk to above the limb. The
dome-like expansion of the faster wavefront (`F') can be seen in these images
like AIA 211 \AA. The faster (`F') and slower (`S') wavefronts can be seen
simultaneously in the image at 19:20:07 UT. To show the propagation of the
faster wavefront, we select a slice cut (indicated by `B') in the plane of the
sky along the wave propagation direction close to the western limb. The
space-time plots of these two slices (A and B) are shown in Figure \ref{st}.
The top panel shows the space-time plot of AIA 211 \AA\ intensity distributions
along slice `A' (refer to image 19:15:12 UT in Figure \ref{aia211}). The top 
panel shows the propagating bright fast and slow wavefronts `F' and `S'. The
coronal dimming behind these fronts is evident in these plots. We measured the
distance-time of these two propagating wavefronts using the top panel. The
measured data points are indicated in the top panel by red (diamond) and blue
(+) for faster and slower wave components. We apply the linear fit to these
data points and attain the speeds of these waves, which are 1086 km s$^{-1}$
and 276 km s$^{-1}$, respectively. The bottom panel shows the fast wave
propagation along slice `B' across the western limb. It shows a diffuse slower
component (marked by `+'), which is the signature of the expanding CME loop.

However, we did not observe any filament or flux rope eruption in AIA EUV images during the flare event. A small loop eruption was observed from the flare site visible in high cadence AIA 94 \AA \ images (Paper I), which moved along the westward direction.
The top panel of Figure \ref{lasco} displays the selected AIA 193 \AA \ running-difference images above the solar western limb across which the wave propagates. The first image at 19:20:07 UT shows the bright circular fast wavefront (indicated by arrows), marked by `F'. In the next images, we can roughly see the shock front straddling over the leading edge of the expanding CME loop. 
For investigating the CME that was associated with the M2.9 flare, we use LASCO C2 \textit{(Large Angle Spectrometric Coronagraph)} white light observation between $\approx$2-6 R$_\odot$
 \cite{brueckner1995}. The bottom panel of Figure \ref{lasco} displays the white light running-difference images, which are combined with AIA 193 \AA \ EUV running-difference images of the same time. These images show the CME
propagation away from the western limb. The second image at 19:24:06 UT shows the expanding CME loop in the AIA field of view, and we see the bright
blob-shaped CME structure. In the coronograph field of view, the CME speed measured from the linear fit is found to be 350 km s$^{-1}$, and it shows an acceleration of 47.5 m s$^{-2}$ during the propagation. The fast shock
disappeared in the LASCO field of view and we could observe only a blob-shaped
structure in the CME (refer to image at 20:36:05 UT). This blob may be linked with the narrow coronal loop, which erupted along with the flare observed in
AIA 94 \AA \ images. 
 %*************************************************************************
\begin{figure}
\centerline{
\includegraphics[width=0.25\textwidth]{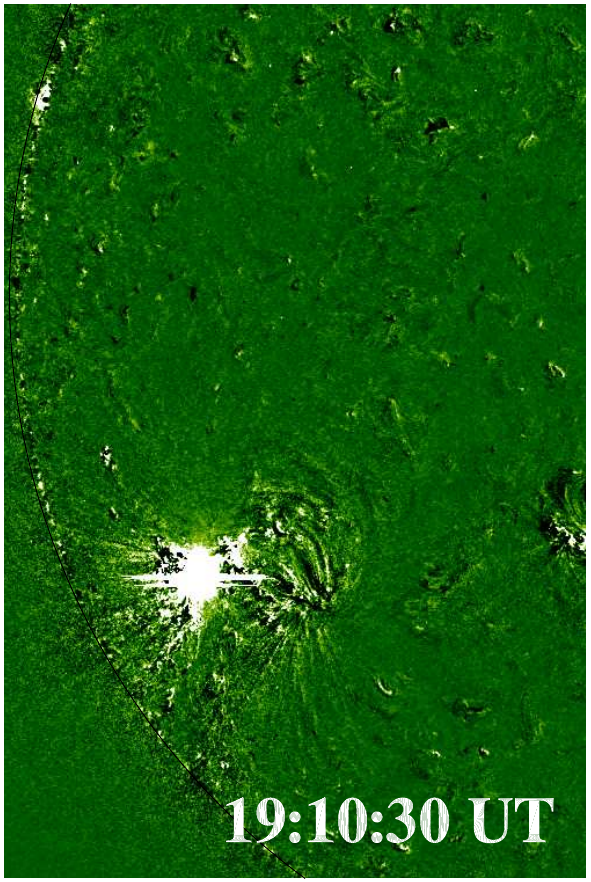}
\includegraphics[width=0.25\textwidth]{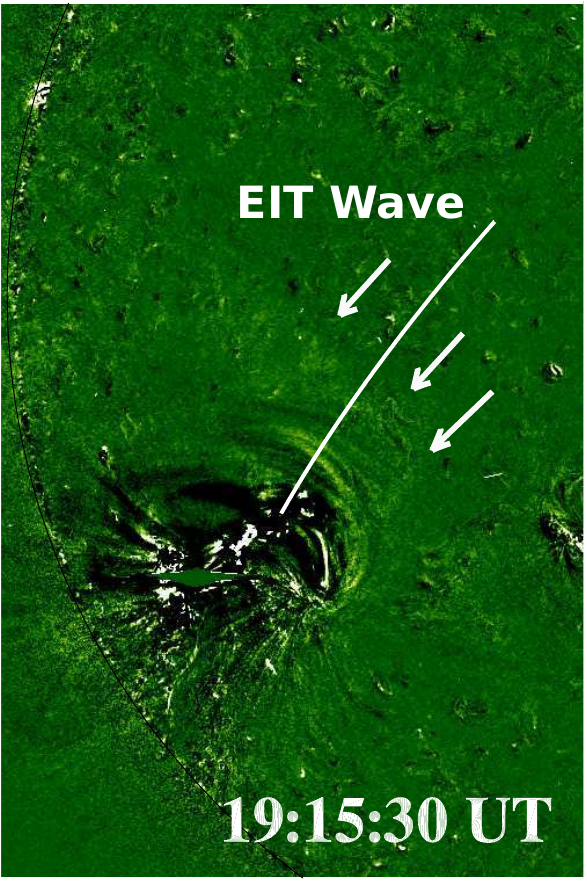}
\includegraphics[width=0.25\textwidth]{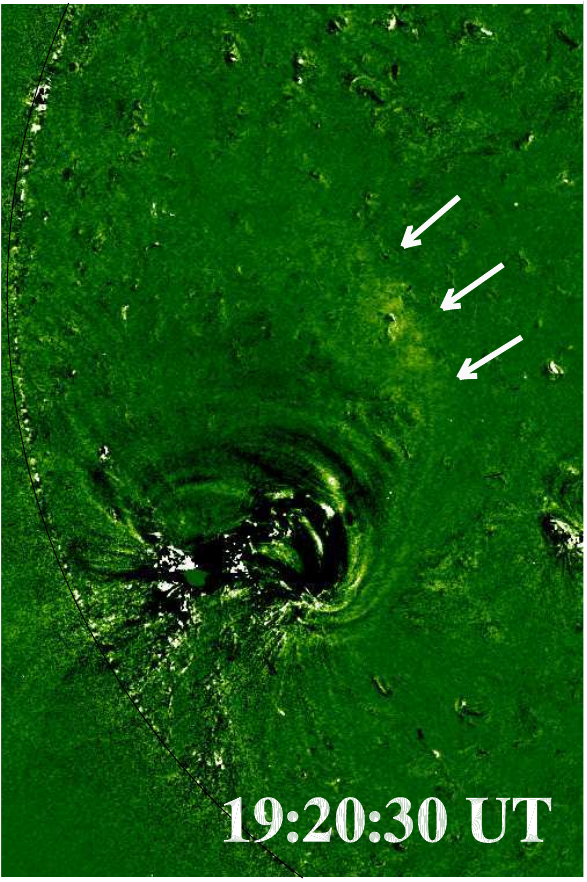}
\includegraphics[width=0.25\textwidth]{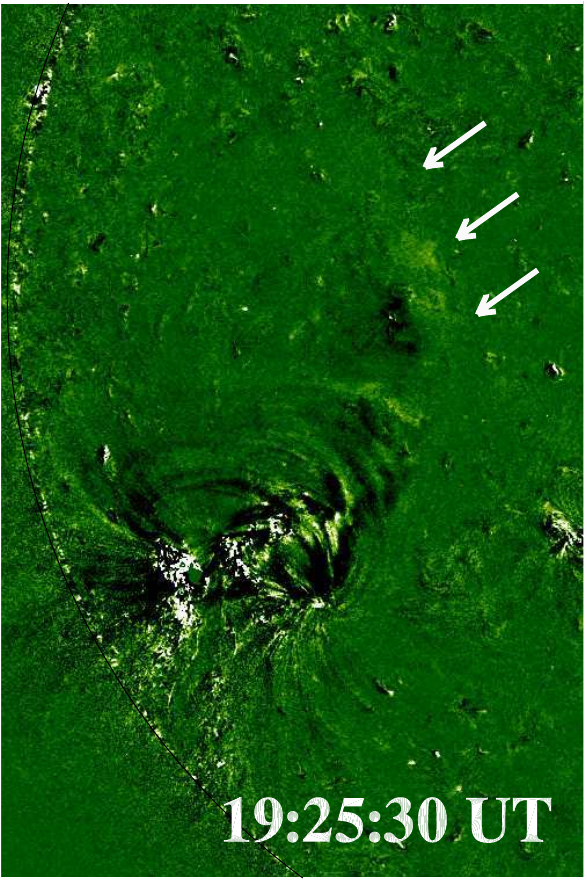}
}
\centerline{
\includegraphics[width=0.25\textwidth]{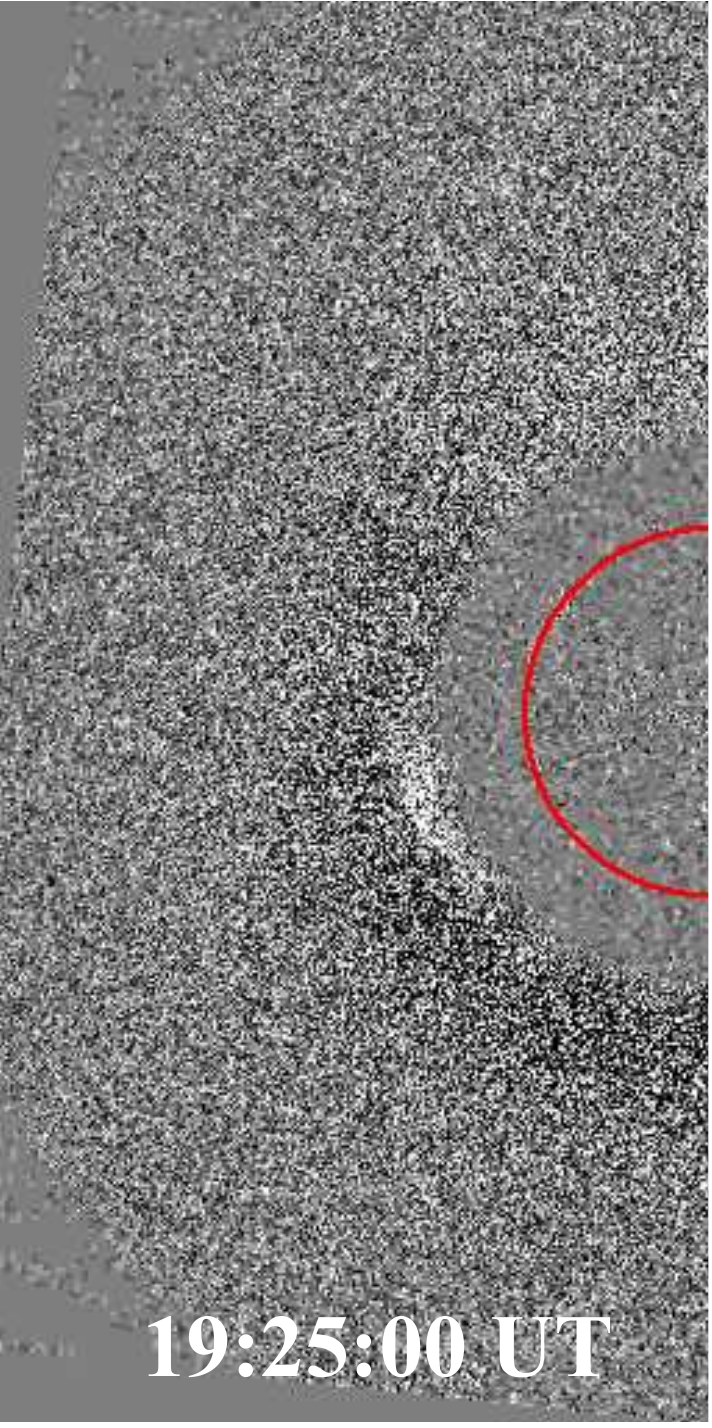}
\includegraphics[width=0.25\textwidth]{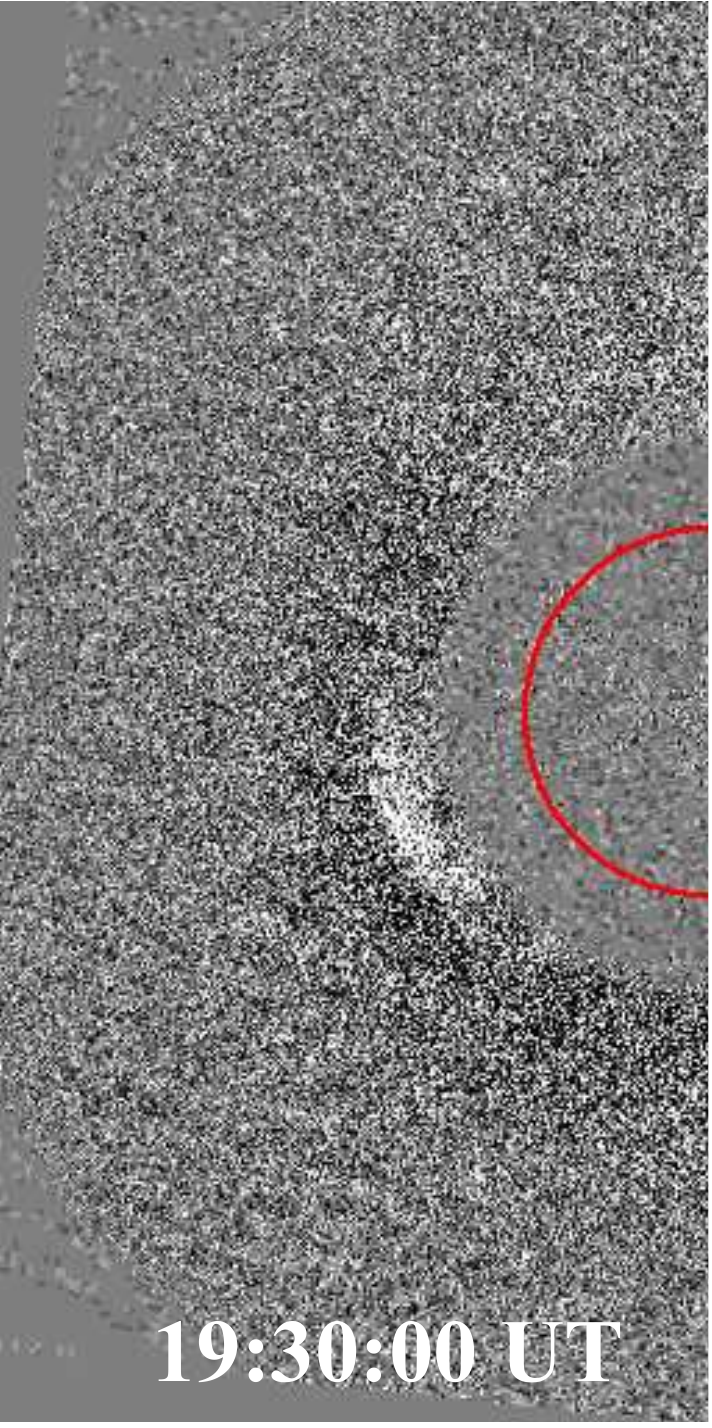}
\includegraphics[width=0.25\textwidth]{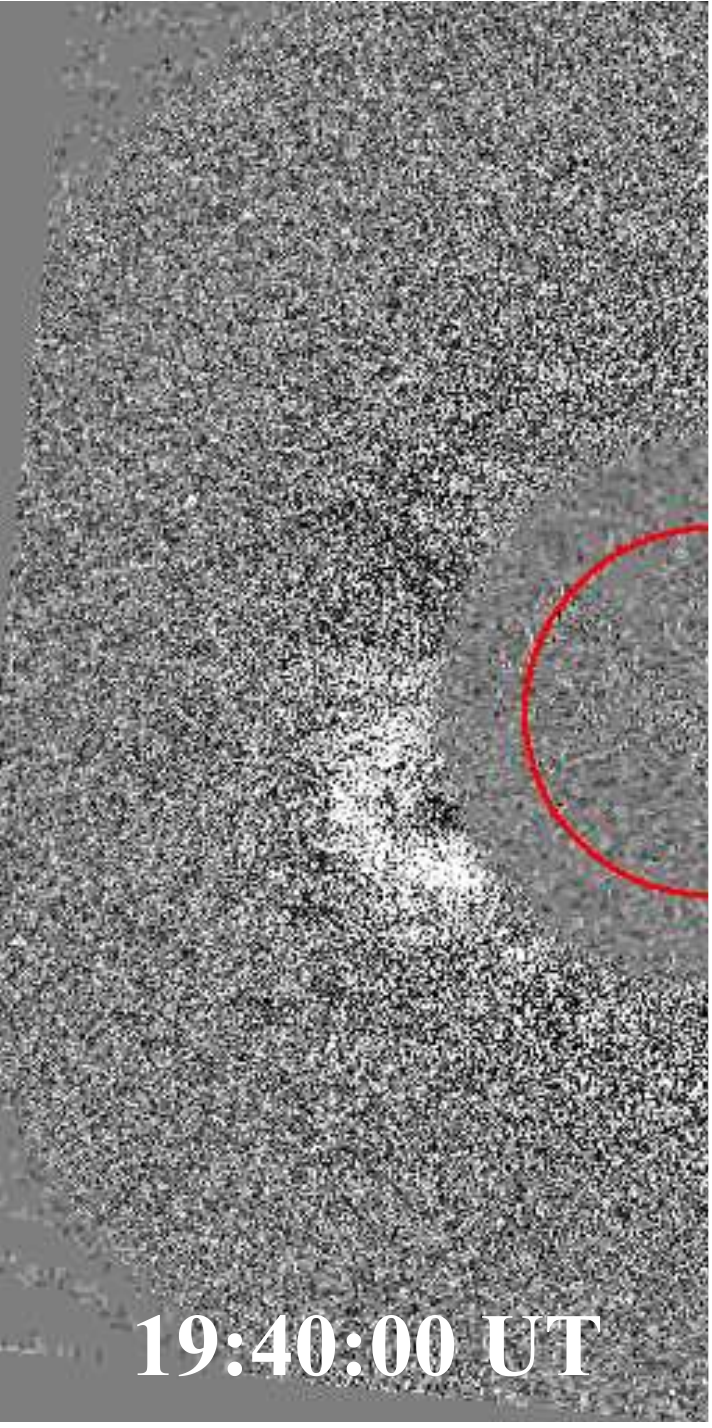}
\includegraphics[width=0.25\textwidth]{fig6f.pdf}
}
\caption{Top: STEREO-A EUV 195 \AA \ running-difference images showing the
flare, dimmings and the propagation of the EIT wave. White line shows the great circle along the solar surface in the direction of wave propagation. Bottom: STEREO-A COR1
base-difference images showing a loop-like slow CME (speed$\approx$340 km
s$^{-1}$).}
\label{stereo}
\end{figure}

%*************************************************************************
 
 We used the STEREO-A ({\it Solar TErrestrial RElations Observatory,} \opencite{kaiser2008}) EUV 195 \AA \ images to see the coronal waves from a different
viewing angle. The size of each image is 2048$\times$ 2048 pixels with a
 1.6$^{\prime\prime}$ per pixel sampling \cite{wuelser2004}. 
 In STEREO-A, the active region was located close to the eastern limb. The top panel of Figure \ref{stereo} shows the 195 \AA \ EUV running-difference images, where we can see a typical EIT wave. We can compare the direction of the EIT wave, which is the same as the slower one seen in the AIA 211 \AA\
base-difference images. In order to estimate the speed of the EIT wave, we have visually tracked the position of the propagating wavefront along the
great circle shown in Figure \ref{stereo}.

The inner coronagraph (COR1) of the
\textit{Sun Earth Connection Coronal and Heliospheric Investigation} (SECCHI, \opencite{Howard2008}) instrument on board STEREO allows us to investigate the CME kinematics in the low corona from 1.4-4.0 R$_{\odot}$ with a high time cadence
$\approx$5 or 10 min and a spatial resolution of 3.75$^{\prime\prime}$.
We used COR1 observations to view the CME during the flare. The bottom panel of Figure 
\ref{stereo} displays the base-difference images of the associated CME observed by COR1. These images confirm a weak and slow CME, which was possibly associated with small loop eruption observed in AIA 94 \AA \ images. The estimated speed of the CME from COR1 height-time measurements was $\approx$340 km s$^{-1}$, which is close to the CME speed measured by LASCO C2 ($\approx$350 km s$^{-1}$). 

%*************************************************************************
\begin{figure}
\centerline{
\includegraphics[width=1.0\textwidth]{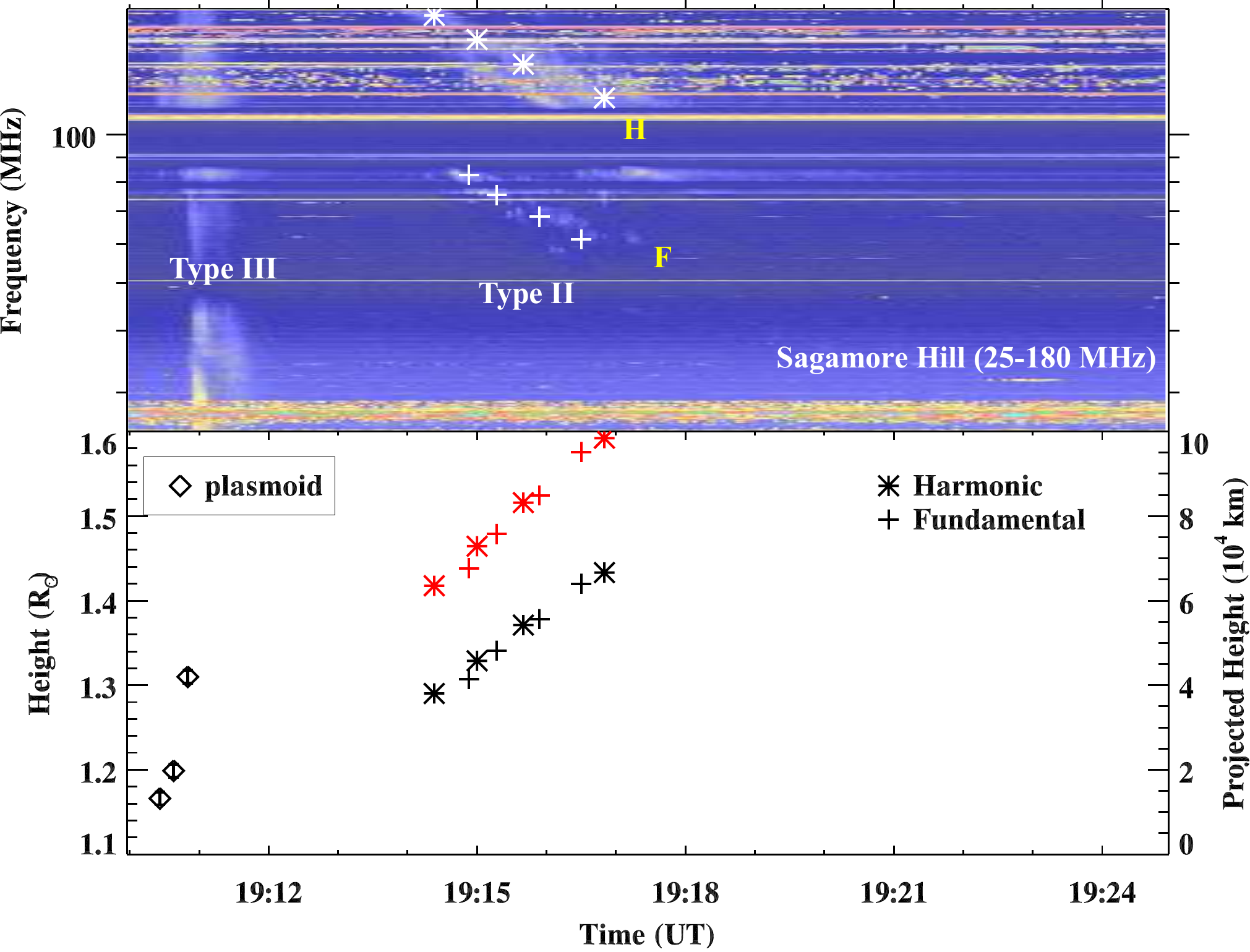}
}
\caption{Top panel: The dynamic radio spectrum in 25-180 MHz observed at
Sagamore Hill station on 16 October 2010, showing type III and type II radio
bursts during the flare. Bottom panel: The source heights of type II burst inferred
from the fundamental band (+) and harmonic band (*) using Newkirk 1-fold (lower) and 2-fold (upper, red) density models, respectively. Projected
height of the plasmoid (diamond) is plotted on the right side of the
$y$-axis.}
\label{type2}
\end{figure}
%%%%%%%%%%%%%%%%%%%%%%%%%%%%%%%%%%%%%%%%%%%%%%%%%%%%%%%%%%%%%%%%%

%*************************************************************************
\begin{figure}
\centerline{
\includegraphics[width=1.0\textwidth]{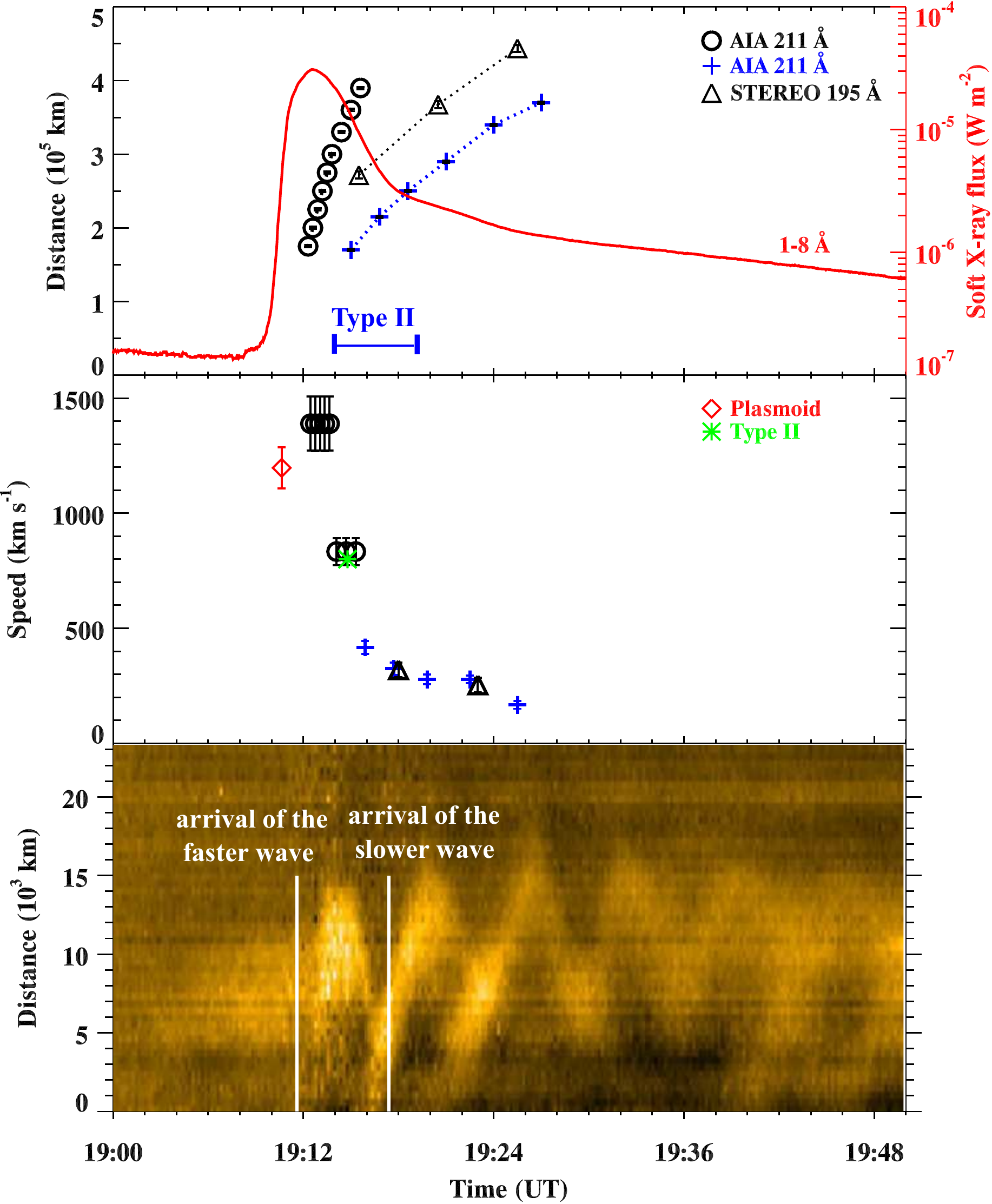}
}
\caption{Top panel: Distance-time profiles of the coronal waves derived
from AIA 211 (circle and plus symbols) and STEREO 195 \AA \ (triangle) images.
GOES soft X-ray profile in 1-8 \AA \ wavelength band is depicted as the red
curve. Duration of the type II radio burst is marked by a blue horizontal line.
Middle panel: Speed profiles for both faster and slower wave. The plasmoid speed (red diamond) 
and coronal shock speed derived from type II (star) radio burst have also been plotted.
Bottom panel: Space-time plot of the oscillating loop along slice cut shown in
the left-panel of Figure \ref{171_slice}.}
\label{loop}
\end{figure}
%%%%%%%%%%%%%%%%%%%%%%%%%%%%%%%%%%%%%%%%%%%%%%%%%%%%%%%%%%%%%%%%%

 The top panel of Figure \ref{type2} displays the dynamic radio spectrum in 25-180 MHz observed at Sagamore Hill radio station, USA \cite{straka1970}. We can see the type III and metric type II radio burst during the flare time. The drifting stripes of metric type II emission ({\it i.e.} fundamental and second harmonic) are known as the signature of coronal shock waves and radio emission frequencies can be converted into emission heights of  the shock by adopting a coronal density model. We used the middle of the emission lane for the fundamental (+) and second harmonic (*) bands. We estimated the shock heights by using one-fold Newkirk coronal density model \cite{newkirk1961}. The corresponding emission heights for both bands have been plotted in bottom panel of Figure \ref{type2}, which shows the emission heights in between 1.3-1.5 R$_\odot$ (from the Sun center).  Using the linear fit to the emission heights, we estimated the shock speed from fundamental and second harmonic, {\it i.e.} $\approx$800 km s$^{-1}$ and 680 km s$^{-1}$ respectively. We also use two-fold Newkirk coronal density model, to estimate the uncertainty caused by (provisional) choice of the density model. The radio-source heights estimated using this model are larger in comparison to previous one (shown by red color). The mean speed of the shock from fundamental and second harmonic are {\it i.e.} $\approx$975 km s$^{-1}$ and 830 km s$^{-1}$ respectively. We also plotted the projected height of the plasmoid (diamond) measured from AIA 94 \AA \ images and the mean speed of the plasmoid was found $\approx$1197 km s$^{-1}$. The plasmoid was observed nearly 3 min prior to the type II radio burst, which may be associated with the formation of shock wave in the corona.

These measurements of the traveling distance of the waves, along with GOES
soft X-ray (1-8 \AA) flux, are plotted in the top panel of Figure \ref{loop}.
The `circle' and `plus' symbols correspond to the AIA 211 \AA \ measurements
whereas `triangle' symbol corresponds to the STEREO 195 \AA.
We measured the position the wave
fronts `F' and `S' at different times using the AIA 211 \AA \ slice `A' in Figure \ref{st}.
The position of the leading edge of the EIT wave shown in STEREO 195 \AA \ images has been measured 
by drawing a great circle from the flare center (indicated in the top panel of Figure \ref{stereo}.)
 The measured speeds of these waves are plotted in the middle panel. 
For AIA observations,
the speed of the faster wave decreases from $\approx$1390 to $\approx$830 km s$^{-1}$,
whereas that of the slower wave decreases from $\approx$416 to $\approx$166 km s$^{-1}$.
In STEREO, the speed of the EIT wave decreased from 320 to 254 km s$^{-1}$. 
The faster wave showed a significant deceleration within the first 5 min. 
The average deceleration of the faster wave is $\approx$--2830 m s$^{-2}$, 
 whereas $\approx$--350 m s$^{-2}$ for slower wave. Note that the uncertainty in the speed estimation is mainly due to the error in the distance measurement in AIA and STEREO, which is taken as 4 pixels ({\it i.e.} 2.4$^{\prime\prime}$ for AIA and 6.4$^{\prime\prime}$ for STEREO).

The speed difference between the EIT wave in STEREO and the faster wave in 
AIA implies the existence of two coronal waves, one faster and another slower
wave, which again confirms our result in Figure \ref{st}. The speed of the
slower wave in AIA is comparable to that observed in STEREO. 
 The faster wave was missed by STEREO. This is probably due to the
low cadence of STEREO, which is not sufficient to detect the faster wave \cite{chen2011}.
Therefore, the EIT wave in STEREO is not cospatial with the fast coronal wave
observed in AIA.  The observational evidence of the coronal Moreton wave ahead
of the EIT wave (using AIA data) was recently confirmed by \inlinecite{chen2011}. 
They found that the speed of the coronal Moreton wave was nearly three times
higher than the EIT wave speed. The present observations also most likely reveal
the existence of the fast-mode MHD coronal Moreton wave ahead of the EIT wave.
In Figure \ref{loop}, we also included the mean speed of the plasmoid (red diamond) and mean speed of the coronal shock (star)
measured from the drift rate of type II (fundamental band), which show the good correspondence between 
all the speeds.

%*****************************************************************************
\begin{figure}
\centerline{
\includegraphics[width=0.8\textwidth]{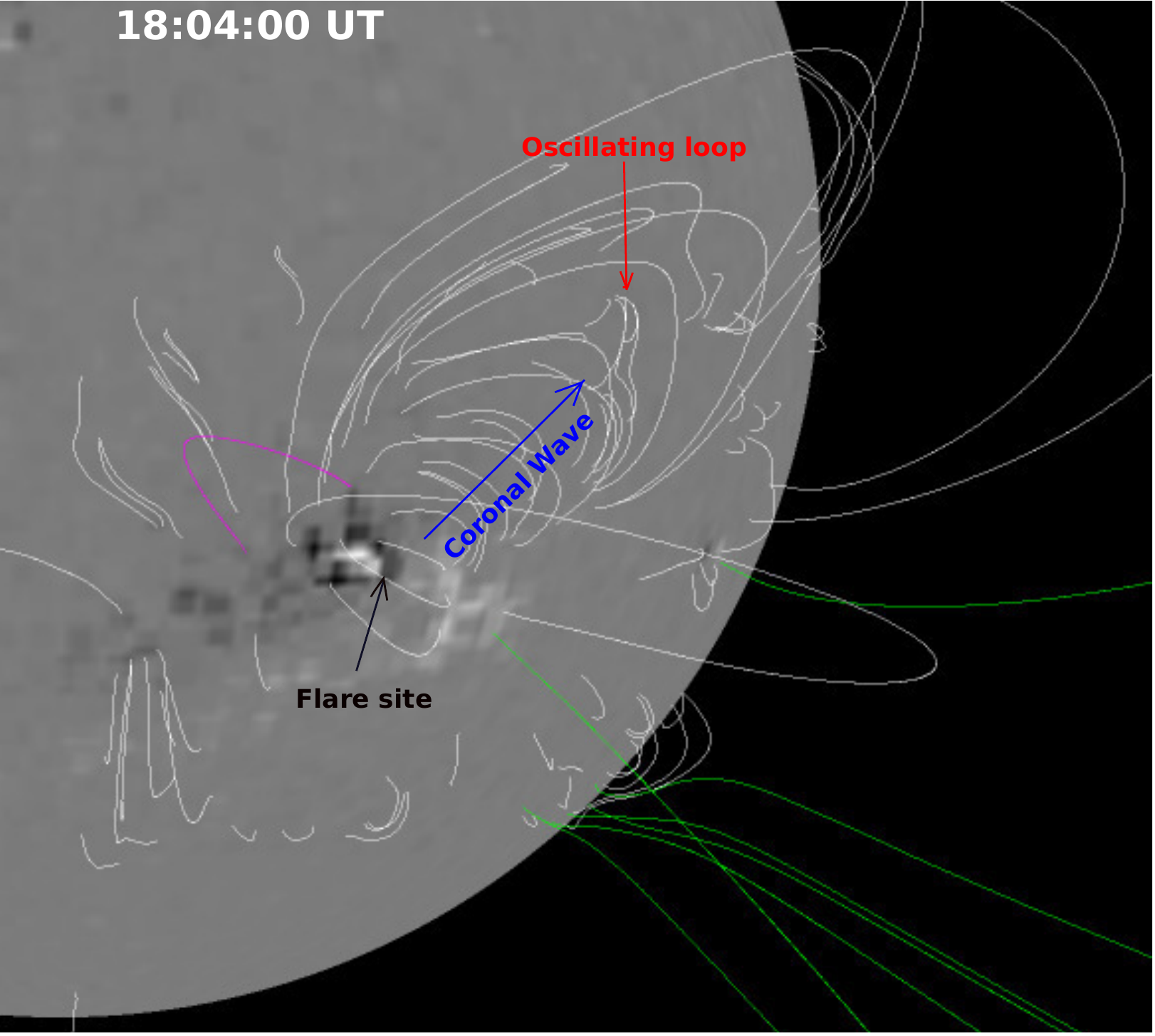}
}
\caption{PFSS extrapolation of the active region NOAA 1112 at 18:04 UT on 16 October 2010.}
\label{pfss}
\end{figure}

%*****************************************************************************
In order to investigate the magnetic field environment of the active region, we used
the potential field source surface (PFSS) extrapolation \cite{alt1969,sch1969}
before the flare event at 18:04 UT. Figure \ref{pfss} shows the PFSS
extrapolation of the active region. The flare site, the coronal wave, and the
oscillating loop are indicated by arrows. Comparing the PFSS extrapolation
with AIA 211 \AA \ images reveals that the field line above the flare site seems to be stretched during the wave propagation. Note that the lateral expansion of the wave/dimming (the direction from the southwest to the
northeast) is negligible. It is probably because the magnetic field lines along the narrow corridor
are closed inside the corridor rather than linking outside, and only these
field lines responded to the CME eruption.

%%************************************************************************
\begin{figure}
\centerline{
\includegraphics[width=0.5\textwidth]{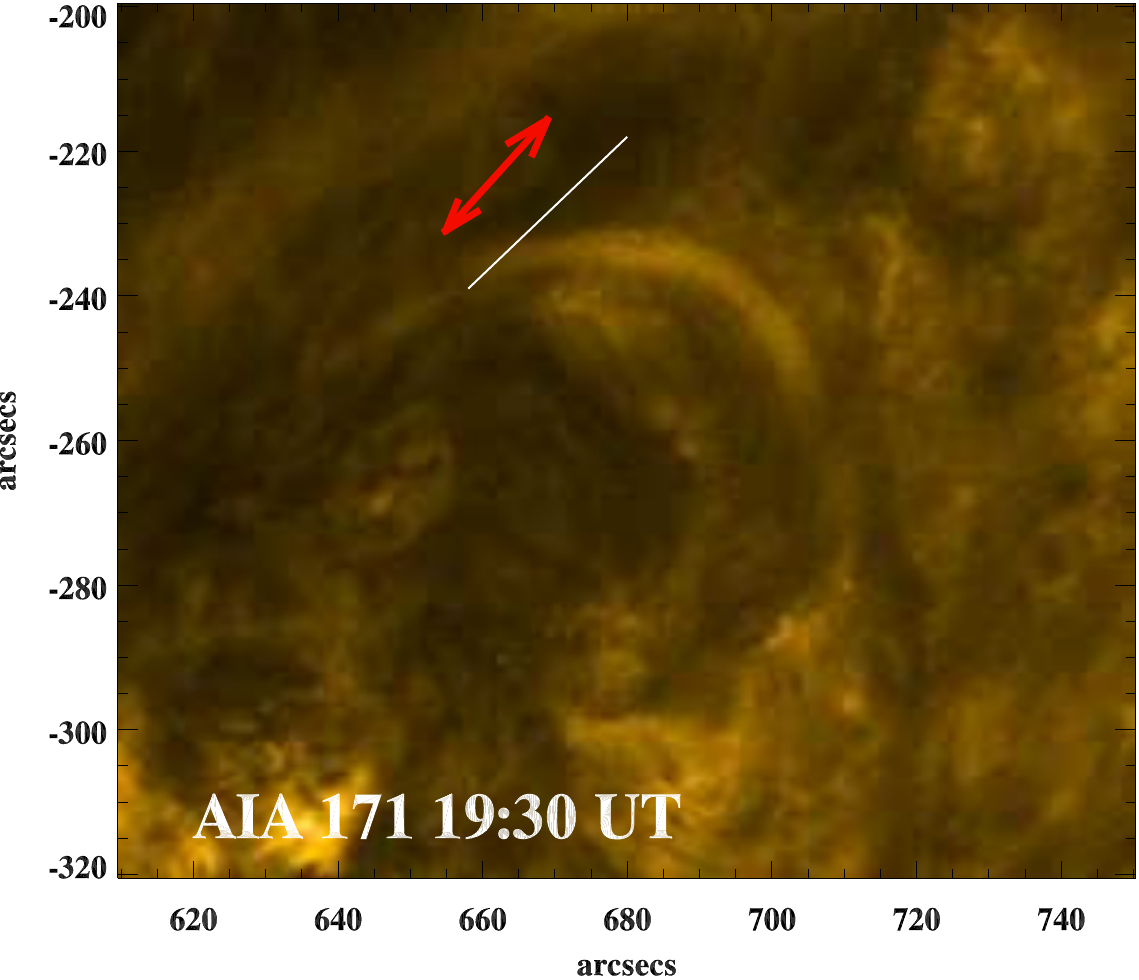}
\includegraphics[width=0.5\textwidth]{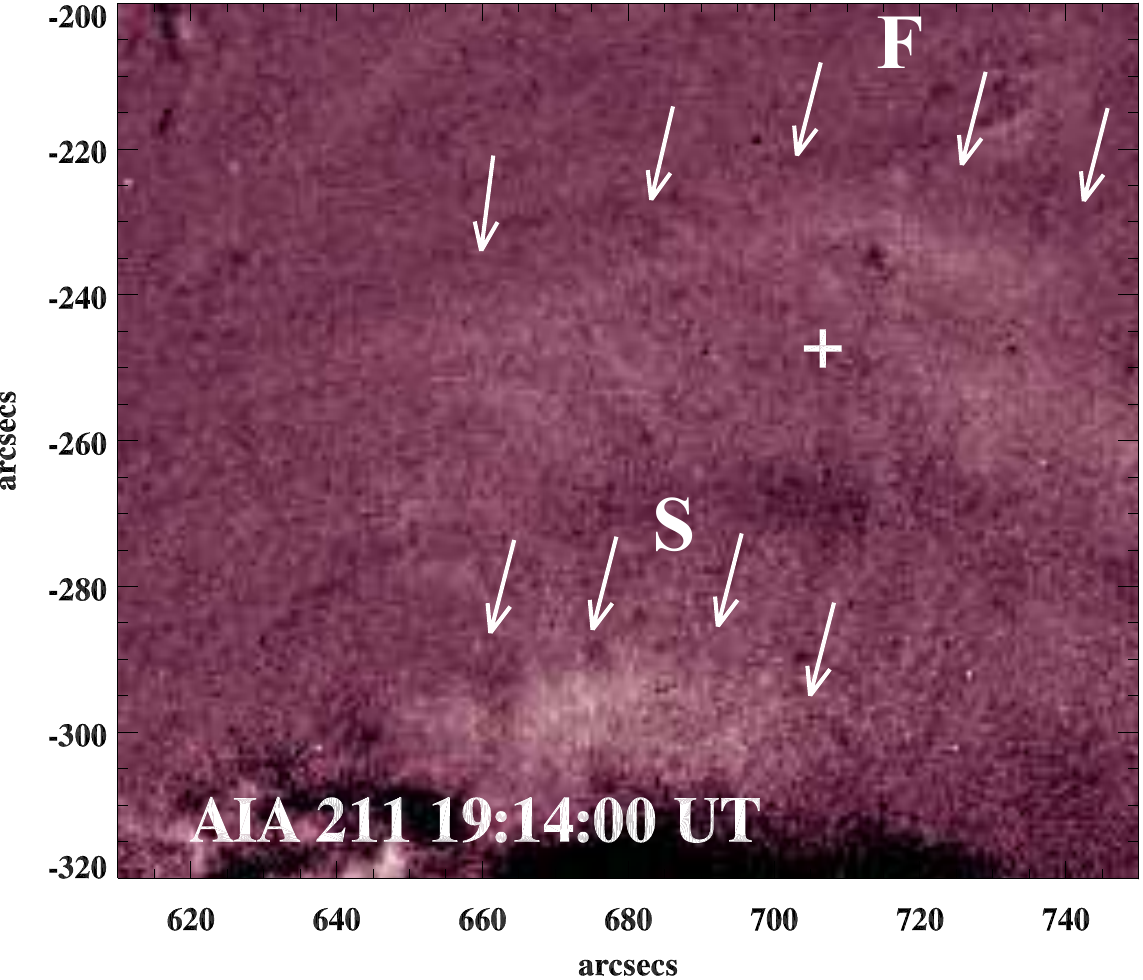}
}
\caption{Left: AIA 171 \AA \ EUV image showing the loop, which presented
transverse oscillations (indicated by the red arrow) during the passage of
coronal waves. The position of the slice cut for space-time plot is marked
as the white line in the image. Right: AIA 211 \AA \ EUV base-difference image showing the
wavefronts (indicated by arrows) of both the faster and the slower wave, marked
by `F' and `S', respectively. The position of the loop apex is marked by the
`+' symbol .}
\label{171_slice}
\end{figure}
%%%%%%%%%%%%%%%%%%%%%%%%%%%%%%%%%%%%%%%%%%%%%%%%%%%%%%%%%%%%%%%%%

\subsection{Coronal Loop-Oscillation}
The right panel of Figure \ref{171_slice} displays the AIA 211 \AA \ EUV
base-difference image at 19:14:00 UT in the enlarged view showing the
wavefronts (indicated by arrows) of both fast and slow waves, which are marked
respectively by `F' and `S'. The position of the loop apex is marked by `+'
symbol. This gives a clear indication of the successive interaction of the faster and slower waves with the coronal loop. The AIA 171 \AA\ image in the
 left panel shows the position of the loop that presented transverse oscillations (indicated by the red arrow, and the online 171 \AA \ movie) during the passage of coronal waves. The location of the slice cut for the
 space-time plot is marked as the white line in this image, and the 
space-time plot is presented in the lower panel of Figure \ref{loop}.

%%%%%%%%%%%%%%%%%%%%%%%%%%%%%%%%%%%%%%%%%%%%%%%%%%%%%%%%%%%%%%%%%%%%%%%%%%%%%%

The loop started oscillating when the leading front of the faster wave
approached it (see aia171.avi, the online movie for loop oscillations). The space-time plot reveals that the amplitude of the
oscillation of selected thread shows an increase before a weak decay, which is very unusual
for coronal loop oscillations \cite{asc2011}. The loop oscillation
continued about five periods and the measured period of oscillation was 6.3 minutes. These measurement of the oscillation period of the loop strand matches well with the findings of \inlinecite{asc2011}. However here, our aim is to shed light on the most likely driver of the coronal loop oscillations. We mark the
arrival times of the faster and slower waves to the coronal loop in the
space-time plot, and it seems that the faster coronal wave is the first driver
to generate the loop oscillations. In addition, type II radio burst was also
observed at the time of coronal wave propagation towards west. The duration of
type II is indicated by a horizontal line in the top panel of Figure 
\ref{loop}. The arrival of propagating faster wave, its relation with the 
flare, the metric type II radio burst, and the starting of loop oscillation, collectively indicate that the coronal loop started to oscillate due to its
interaction with a coronal shock wave. 

The peculiar behavior of this loop oscillation event is that the
oscillation did not show strong decay as usual, and instead, its
amplitude was increasing in the first two periods. In this sense, it should be
noted that the slower wave (EIT wave) passed through the location of coronal 
loop after the faster wave. The EIT wave arrived at the loop at about 19:16:24
UT (refer to Figure \ref{loop}). Coincidentally, the amplitude of the 
oscillation increased after the arrival of the EIT wave (front S) and became
stronger (bottom panel of Figure \ref{loop}). The loop oscillation was
observed till the wave passed there. Later when it moved out, loop
oscillations decayed weakly. Therefore, it is likely
that the passage of the slower wave caused the stronger loop oscillation for a
longer time. 

\inlinecite{eto2002} also have found that the filament winking was initiated by the
passage of a Moreton wave, and was enhanced when the EIT wave passed the
filament. In their study, the times of visibility for the Moreton wave did not
overlap with those of the EIT wave. Instead, the continuation of the Moreton wave was
implied by the filament oscillation. Using the position and speed measurements, they clearly showed that the Moreton wave differed physically
from the EIT wave in their case. In our case, the loop oscillation behavior 
generated by the impact of the faster and the slower coronal waves are
consistent with the filament oscillations initiated by the passage of the
Moreton and EIT waves. 
 Recently, \inlinecite{asai2012} have reported the first
   simultaneous observation of an H$\alpha$ Moreton wave,
   the corresponding EUV fast coronal waves, and a slow
   and bright EIT wave. They also observed the filament/prominence oscillations when the wave approached it. However, we do not have the H$\alpha$ observations during the flare. But, we see the propagating brightening/disturbance in AIA 304 \AA \ images (correspond to chromosphere and transition region) in the similar direction of the EUV wave, visible in AIA 211 and 193 \AA \ images. Our observational findings are also consistent with \inlinecite{asai2012}.

%%%%%%%%%%%%%%%%%%%%%%%%%%%%%%%%%%%%%%%%%%%%%%%%%%%%%%%%%%%%%%%%%%%%%%%%%%%%%%

\section{Discussion and Conclusions}
We analyzed the multiwavelength observations of the M2.9/1N flare occurred
on 16 October 2010 from AR NOAA 1112. We first discuss the identification of
two waves associated with the flare/CME event, {\it i.e.,} a faster coronal Moreton
wave and a slower EIT wave. According to the SDO/AIA observations, the flare
and the CME were associated with a faster and a slower waves, which moved
towards the west decelerating from $\approx$1390 to $\approx$830 km s$^{-1}$ and
 from $\approx$416 to $\approx$166 km s$^{-1}$, respectively. In STEREO 195 \AA \, only
one diffuse EIT wave was discernable, decelerating from $\approx$320 to $\approx$254 km s$^{-1}$. The slower wave in SDO/AIA is interpreted as a
classical EIT wave, consistent with the STEREO observations.

According to Uchida's Model (\citeyear{uchida1968}, \citeyear{uchida1970}), 
the Moreton wave is a sweeping skirt on the chromosphere of the MHD fast-mode
shock wave which propagates in the corona. Therefore, this model predicts the
existence of a coronal counterpart of the chromospheric Moreton wave at the
same place and with the same velocity as that of the Moreton wave.
\inlinecite{thompson2000} reported that there are two components in EIT waves,
{\it i.e.,} bright/sharp and diffuse EIT waves. The sharp EIT wave and H$\alpha$
Moreton wave are cospatial, whereas the relationship between the diffuse EIT
and Moreton wave was not clear. In the present event, we observed the diffuse EIT wave in STEREO 195 \AA\ images, which was cospatial with the slower wave
observed by SDO/AIA. \inlinecite{wu01} and 
\inlinecite{warmuth2001} suggested that the diffuse EIT waves are decelerated
Moreton waves, namely that not only the sharp EIT waves but also the diffuse
EIT waves are coronal counterparts of the chromospheric Moreton waves. 
However, \inlinecite{eto2002} found that the diffuse EIT wave was not the coronal counterpart of the chromospheric Moreton wave in their analysed event. 
 In the present paper, we revealed the existence of both faster and slower wavefronts, which are not co-spatial. In addition, they have very different
velocities. The existence of these two waves is consistent as predicted in  \inlinecite{chen2002} model and it was confirmed by \inlinecite{harr03} and
\inlinecite{chen2011}.
 
 A remote small
coronal loop started to oscillate with a period of 6.3 minutes as the faster
wave hit it. The detailed study of the loop oscillation has been
presented by \inlinecite{asc2011}. We suggest that the faster wave is most
likely the first driver of loop oscillation, and the oscillation was enhanced
by the ensuing EIT wave. PFSS extrapolation and the direction of the fast wavefront in association with loop oscillation suggest that the coronal Moreton wave propagates across the closed magnetic loops. The visibility of the coronal Moreton wave may be related to the local magnetic field, and tends
to be enhanced at weaker magnetic field \cite{uchida1970}. 

 The initiation of a filament oscillation that preceded the arrival of the
EIT wave in \inlinecite{eto2002} was suggested as evidence in support of the
idea that EIT waves are not coronal Moreton waves. On the other hand,
\inlinecite{warmuth2004} interpreted this event in terms of a tilted coronal wavefront: since the filament is located higher up, the more tenuous upper --
and thus less observable--parts of the wavefront will reach it first.
Furthermore, determining at which time the filament actually begins to
oscillate can be quite ambiguous, so that the possible errors can be much
larger than the errors on the wavefronts \cite{warmuth2010}. But, in the
present case, we have shown that the triggering of loop oscillation at the
arrival of the faster wavefront, and that the oscillation was enhanced with 
the passage of the slower wavefront, {\it i.e.} the EIT wave.

 A metric type II burst was also observed during the propagation of coronal
waves. The speed of the shock wave derived from type II frequency drift rate 
($\approx$800 km s$^{-1}$) matches well with the speed of the faster coronal
wave associated with the flare/CME. The type II radio burst may be associated
with the high speed coronal wave as it moves nearly with Alfv\'enic speed in
the corona.  The observed faster coronal wave is probably the
fast-mode coronal Moreton wave, and the presence of type II during this time
supports the presence of a fast-mode MHD shock wave as predicted by
\inlinecite{uchida1974}. We found good temporal, spatial and speed matching between EUV coronal Moreton wave and the shock wave derived from type II radio burst.

 In a statistical analysis of coronal loop oscillations
observed by TRACE, \inlinecite{hudson2004} showed the strong association of
TRACE loop oscillation events with type II bursts indicating that some of them
were directly caused by blast waves. In their observations, only certain loops
oscillate, whereas other nearby loops remain stationary, which was consistent
with the highly directional nature of blast waves \cite{smith1971,warmuth2004a}. On the other hand, a piston-driven-type shock could be launched
by ejecta with a smaller scale ({\it e.g.,} sprays or ejecta observed with the
Yokhoh soft X-ray telescope instead of an initial pressure pulse. For example,
\inlinecite{klein1999} have shown X-ray blob (projected speed$\approx$770 km
s$^{-1}$) as a plausible driver of a fast shock in the corona.
 They could act as a temporary piston, and either they could generate a perturbation that then steepens into a shock or there could be a short phase of a driven shock, after which the shock propagates freely \cite{warmuth2004,veronig2010,muhr2011}. In the present case, we observe (in AIA 94 \AA \ images) a high speed plasmoid (projected speed$\approx$1197 km s$^{-1}$) moving away from the flare site during the flare impulsive phase (Paper I) and this eruption may be responsible to drive high speed shock and type II observed in this event. This favors the scenario for the piston driven shock.

 Recently, \inlinecite{warmuth2011} analysed large sample of 176 EIT wave events and based on their kinematical behavior, they found the evidence for three distinct populations of coronal EUV waves: initially fast waves (v$\geq$ 320 km s$^{-1}$ ) that show pronounced deceleration (class 1 events), waves with moderate (v$\approx$170-320 km s$^{-1}$ ) and nearly
 constant speeds (class 2), and slow waves (v$\leq$130 km s$^{-1}$ ) showing a rather erratic behavior (class 3). They explained class 1 and 2 in terms of the fast mode wave/shock model, whereas class 3 events due to the magnetic reconfiguration. By combining data from AIA and EIS, \inlinecite{harra2011} and \inlinecite{veronig2011} examined
a coronal wave and found that the main wave front travels at $\approx$500 km s$^{-1}$
and is strongly redshifted ({\it i.e.,} as the wave propagates it also
pushes plasma downward with a speed of
$\approx$20 km s$^{-1}$ ). They concluded that the observed wave was
generated by the outgoing CME, as in the scenario for the classic Moreton wave ({\it i.e.} fast MHD wave), which pushes down the chromospheric plasma along the shock front.

Our observations reveal the signature of fast coronal Moreton wave and associated loop oscillation that was initiated by its interaction. On the other hand, the slower wave observed in this event cannot be the top part of the CME leading loop since it impacted the small coronal loop. It could be associated with the leg of the CME leading loop, while EIT wave was already found to be cospatial with the CME leg \cite{chen2009,dai2010}. Therefore, we conclude that the slower wave is the classical EIT wave.

In conclusion, we presented the multiwavelength observations of both the
faster and slower coronal waves, which may be the first and the second drivers
of the oscillations of a remote loop. Using the high spatial and temporal data
from space and ground based instruments, further studies should be performed
in order to shed more light on the flare processes and their association with
large-scale coronal waves.  

%%%%%%%%%%%%%%%%%%%%%%%%%%%%%%%%%%%%%%%%%%%%%%%%%%%%%%%%%%%%%%%%%%%%%%%%%%%
\begin{acks}

We express our gratitude to the referee for his/her valuable and constructive
comments/suggestions which improved the manuscript considerably. SDO is a
mission for NASA's Living With a Star (LWS) Program. We thank the STEREO/SECCHI teams for their open
data policy.
 We are thankful for the radio data obtained from Sagamore Hill station. SOHO is a project of international cooperation between ESA and NASA. PFC is supported by the Chinese foundation NSFC
(11025314, 10878002, and 10933003) and 2011CB811402. PK thanks to Prof. D.E.
Innes for several fruitful discussions during his visit to MPS. PK thanks to Dr. A.K. Srivastava for reading/discussing the manuscript. This work has been supported by the ``Development of
Korea Space Weather Center" project of KASI, and the KASI basic research fund.
 \end{acks}
%%%%%%%%%%%%%%%%%%%%%%%%%%%%%%%%%%%%%%%%%%%%%%%%%%%%%%%%%%%%%%%%%%%%%%%%%%%%%
\bibliographystyle{spr-mp-sola}
\bibliography{reference}  
\end{article} 
\end{document}